\documentclass[sigplan,screen]{acmart}

\newif\ifcopyright
\copyrightfalse

\ifcopyright
\setcopyright{acmcopyright}
\else
\setcopyright{none}
\fi

\acmPrice{15.00}
\acmDOI{10.1145/3314221.3314634}
\acmYear{2019}
\copyrightyear{2019}
\acmISBN{978-1-4503-6712-7/19/06}
\acmConference[PLDI '19]{Proceedings of the 40th ACM SIGPLAN Conference on Programming Language Design and Implementation}{June 22--26, 2019}{Phoenix, AZ, USA}
\acmBooktitle{Proceedings of the 40th ACM SIGPLAN Conference on Programming Language Design and Implementation (PLDI '19), June 22--26, 2019, Phoenix, AZ, USA}

\usepackage{url}
\usepackage{booktabs}
\usepackage{listings}
\usepackage{proof}
\usepackage{mathrsfs}
\usepackage{mathtools}
\usepackage{amssymb,amsmath,amsthm, amsfonts}
\usepackage{algorithm}
\usepackage[noend]{algpseudocode}
\usepackage{tikz}
\usetikzlibrary{calc,shapes.multipart,chains,arrows,fit,shapes.misc}
\tikzstyle{node}=[
  rectangle split,
  rectangle split parts=2,
  draw, text centered]

\newcommand{\code}[1]{\texttt{#1}}
\newcommand{\tool}{\textsf{SLING}}
\newtheorem{definition}{Definition}

\newenvironment{algosummary}{
  \begin{flushleft}}
  {\end{flushleft}}


\newcommand{\Keyword}[1]{\ensuremath{\mathrel{\mathbf{#1}}}}
\newcommand{\Summary}[1]{\textbf{#1}}
\newcommand{\Assign}[2]{\ensuremath{#1}~\ensuremath{\leftarrow}~\ensuremath{#2}}
\algdef{Se}[IF]{If}{EndIf}[1]{\algorithmicif\ \ensuremath{#1}\ \algorithmicthen}{}%
\algdef{Se}[IF]{ElsIf}{EndIf}[1]{\algorithmicelse\ \algorithmicif\
  \ensuremath{#1}\ \algorithmicthen}{}%
\algdef{Se}[FOR]{For}{EndFor}[1]{\algorithmicfor\ \ensuremath{#1}\ \algorithmicdo}{}%
\algdef{Se}[WHILE]{While}{EndWhile}[1]{\algorithmicwhile\ \ensuremath{#1}\ \algorithmicdo}{}%
\renewcommand\Return[1]{\Keyword{return} \ensuremath{#1}}
\renewcommand{\Call}[2]{\proc{#1}\ensuremath{(#2)}}

\newcommand{\proc}[1]{\ensuremath{\mathsf{#1}}}

\newcommand{\pred}[1]{\ensuremath{\mathsf{#1}}}
\newcommand{\val}[1]{\ensuremath{\mathsf{#1}}}

\newcommand{\formPtr}[3]{\ensuremath{{#2}{
    \overset{\mkern-2mu\hspace{-1pt}#1}\mapsto}{#3}}}
\newcommand{\formPred}[2]{\ensuremath{{#1}{(#2)}}}

\def\definedbnf{\mathrel{\Coloneqq}}

\def\emp{\mathord{\pred{emp}}}
\def\nil{\mathord{\val{nil}}}

\def\true{\mathord{\val{true}}}

\def\sort{\tau}
\def\inst{\iota}
\def\P{\mathord{\pred{p}}}

\def\cmul{\mathrel{\mkern1mu\cdot\mkern1mu}}

\def\sep{*}

\def\satisfies{\mathrel{
    \raisebox{0.1em}{\scalebox{1}[0.85]{\ensuremath{|}}}
    \mkern-2.5mu\mkern-1mu
    \raisebox{0.08em}{\scalebox{0.95}[1]{\ensuremath{=}}}}}

\def\reduces{\mathrel{
    \raisebox{0.1em}{\scalebox{1}[0.85]{\ensuremath{\parallel}}}
    \mkern-3.2mu
    \raisebox{0.1em}{\scalebox{0.9}[1]{\ensuremath{-}}}}}

\def\subtype{\mathrel{<:}}

\def\DefRightarrow{\mathbin{\stackrel{
    \mathclap{\mbox{$\scriptscriptstyle \mathrm{def}$}}}{
    \Rightarrow}}}

\def\setempty{\ensuremath{\mathord{\varnothing}}}
\def\hdisjoins{\mathrel{\#}}
\def\hunions{\circ}
\def\Val{\mathord{\code{Val}}}

\newcommand{\seq}[2]{\ensuremath{(#1)_{#2}}}

\newcommand{\fDom}[1]{{\ensuremath{\mathrm{dom}(#1)}}}
\newcommand{\evalVar}[2]{\ensuremath{{#2}({#1})}}
\newcommand{\evalForm}[2]{\ensuremath{\llbracket {#1} \rrbracket_{#2}}}
\newcommand{\modelExt}[3]{\ensuremath{[{#1}\,|\,#2\,{:}\,\,#3]}}
\newcommand{\modelcheck}[3]{\ensuremath{{#1} \reduces {#2}
    \rightsquigarrow {#3}}}
\newcommand{\model}[3][]{\ensuremath{{#2}
    \ifthenelse{\equal{#1}{}}{\satisfies}{\satisfies_{#1}} {#3}}}

\def\definedas{\mathrel{\stackrel{\makebox[0pt]{\mbox{\normalfont\tiny def}}}{=}}}
\def\Dll{\mathsf{\pred{dll}}}

\def\res{\mathord{\code{res}}}

\begin{document}
\title[SLING: Using Dynamic Analysis to Infer Program Invariants $\ldots$]{SLING: Using Dynamic Analysis to Infer Program Invariants in Separation Logic}

\author{Ton Chanh Le}
\affiliation{%
  \institution{Stevens Institute of Technology}
  \city{Hoboken}
  \state{New Jersey}
  \country{USA}
}
\email{letonchanh@gmail.com}

\author{Guolong Zheng}
\affiliation{
  \institution{University of Nebraska-Lincoln}
  \city{Lincoln}
  \state{Nebraska}
  \country{USA}
}
\email{gzheng@cse.unl.edu}

\author{ThanhVu Nguyen}
\affiliation{
  \institution{University of Nebraska-Lincoln}
  \city{Lincoln}
  \state{Nebraska}
  \country{USA}
}
\email{tnguyen@cse.unl.edu}

\begin{abstract}
We introduce a new dynamic analysis technique to discover invariants in separation logic for heap-manipulating programs.
First, we use a debugger to obtain rich program execution traces at locations of interest on sample inputs.
These traces consist of heap and stack information of variables that point to dynamically allocated data structures.
Next, we iteratively analyze separate memory regions related to each pointer variable and search for a formula over predefined heap predicates in separation logic to model these regions.
Finally, we combine the computed formulae into an invariant that describes the shape of explored memory regions.

We present {\tool}, a tool that implements these ideas to automatically generate invariants in separation logic at arbitrary locations in C programs, e.g., program pre and postconditions and loop invariants.
Preliminary results on existing benchmarks show that {\tool} can efficiently generate correct and useful invariants for
programs that manipulate a wide variety of complex data structures.

\end{abstract}



\maketitle

\emergencystretch 1em

\section{Introduction}
\label{sec:intro}

A program invariant is a property that holds whenever program execution reaches a specific location.
For example, a loop invariant can indicate a relation among the program variables at the loop entrance.
Invariants help prove program correctness, e.g., classical verification approaches by Floyd-Hoare and Dijkstra~\cite{hoare:69:axiomatic,dijkstra:cacm75:wp} can be automated when given needed loop invariants and the infamous Heartbleed bug can be avoided by preserving an invariant capturing the proper size of the received payload message~\cite{fava:lpar15:gamifying}.
Invariants also help developers understand programs, e.g., showing interesting or unexpected behaviors, and even discover non-functional bugs, e.g., revealing that the program has an unusual high runtime complexity~\cite{nguyen:ase17:syminfer}.
Invariants are also useful in other programming tasks, including documentation, maintenance, code optimization, fault localization, program repair, and security analysis~\cite{leroy:popl06:,ball:ifm04:slam,flanagan:fm01:houdini,ernst:scp07:daikon,le:issta16:faultlocinvs,perkins:sosp09:clearview,nguyen:ase17:syminfer}.

Unfortunately, software developers appear to perceive a ``specification burden''~\cite{ball:ifm04:slam} which leads them to eschew the writing of invariants in favor of executable code.
For the past decade, researchers have been chipping away at this challenge of automatic invariant generation using static or dynamic analyses.
A static analysis can reason about all program paths soundly, but doing so is
expensive and is only possible to relatively small programs or simple forms of invariants, e.g.,
simple list structures~\cite{berdine:fsttcs04,cook:concur11,perez:pldi11}.
In contrast, dynamic analysis focuses on program traces observed from running the
program on small sample inputs, and thus provides no correctness guarantee on
generated invariants. However, dynamic analysis is generally efficient and can infer
expressive invariants because it only analyzes a finite, typically small, set of
traces.

Existing invariant techniques often focus on invariants over scalar variables, e.g., relations among numerical values~\cite{padhi:pldi16:pie,garg:popl16:ice,nguyen:ase17:syminfer}.
However, modern programs construct and manipulate data structures, i.e., highly-structured sets of memory locations within which these scalar values are stored.
Examples of such data structures are dynamically-allocated objects, e.g., heap-based objects created via the \code{new} keyword, standard data structures, e.g., lists and trees, or customized and user-defined structures that extend the standard ones and contain other structures internally.
Understanding and reasoning about these heap-based programs are more challenging, e.g., even the task of accessing a variable requires checking if it points to a valid memory region (to avoid null pointer dereferencing).

An emerging approach to analyzing heap programs is to use invariants written in \emph{separation logic} (SL) to represent memory structures~\cite{reynolds:lics02:sl,ohearn:csl01:sl}.
SL extends classical logic and allows for compact and precise representations of program semantics and reasoning to be localized to small portions of memory.
In the last decade, research in SL has grown rapidly and led to practical techniques used in tools such as the Facebook Infer (FBInfer) analyzer~\cite{fb:web:fbinfer}.

Most existing SL works focus on static analyses to obtain sound results, and therefore can only consider simple classes of invariants or programs, e.g., to support the goal of ``move fast to fix more things''~\cite{ohearn:web16:,calcagno:nasafm:moving}, FBInfer only considers simple data structures  and restricts supported language features.
Moreover, while many static analyzers, including FBInfer, compute SL invariants internally to verify programs, we are aware of only a few researchers who have investigated reifying those invariants for consumption by developers, and even then only for a restricted language of list manipulating~\cite{magill:space06:inferring} or tree traversing programs~\cite{brockschmidt:sas17:locust}.
Also, most static SL tools aim to infer sufficiently strong invariants to achieve a specific goal, e.g., to prove memory safety or (programmer-provided) postconditions, and thus are not well suited for discovering useful invariants to help understand code that lacks such formal specifications.

In this work, we introduce {\tool} (\textbf{S}eparation \textbf{L}ogic \textbf{In}variant \textbf{G}eneration), a tool that dynamically discovers SL invariants for heap programs.
{\tool} takes as inputs a program, a location of interest, a set of predefined predicates defining data structures, and a set of sample inputs.
{\tool} next runs the program on the inputs and uses a debugger to obtain traces capturing memory information of the variables at the considered location.
These traces consist of the contents of the stack and heap of the program.
{\tool} then iteratively analyzes variables using these traces to compute invariants.
For each pointer variable, {\tool} generates SL formulae using predefined predicates to model the traces describing memory regions related to the variable.
{\tool} also propagates computed information to improve the analysis of other variables in subsequent iterations.
Finally, {\tool} combines the obtained formulae into a final invariant that represents the explored memory regions.

We use {\tool} to infer invariants for 157 C programs taken from two existing benchmarks~\cite{web:vcdryad} and~\cite{brotherston:sas14:caber}.
These programs implement basic algorithms over standard data structures (e.g., singly-linked, doubly-linked, circular lists, binary trees, AVL, red-black trees, heaps, queues, stacks, iterators) and complex functions from open source libraries and the Linux kernel that manipulate customized data structures.
Preliminary results show that {\tool} can efficiently generate invariants that are correct and more precise than the documented invariants and specifications in these programs.
Even when given incomplete traces, the tool can still discover partial invariants that are useful for users.
We also show that {\tool}'s invariants can help reason about nontrivial bugs and reveal false positives in modern SL static analyzers.
We believe that {\tool} strikes a practical balance between correctness and expressive power, allowing it to discover complex, yet interesting and useful invariants out of the reach of the current state of the art.

\lstset{
  basicstyle=\ttfamily\footnotesize,
  language=C,
  keywordstyle=\color{blue},
  mathescape,
  stepnumber=1,
  numbers=left,
  numbersep=8pt,
  backgroundcolor=\color{white},
  tabsize=4,
  showspaces=false,
  showstringspaces=false,
  emph={},
  emphstyle=\color{red}\bfseries,
  breaklines=true
}

\newcommand{\pretraces}[1]{
\begin{tikzpicture}[scale=#1,transform shape]
  \tikzstyle{part} = [rectangle, draw=black!50, fill=black!20, inner sep=0.5pt, outer sep=0pt,minimum size=1em]

  \node[part] (A1) {};
  \node[part] (A2) [right= 0pt of A1, label={[name=a1] right:\scriptsize{$\code{0x01}$}}] {};
  \node (A0) [above= 0pt of A1]{};

  \draw (A2.south west) -- (A2.north east);
  \draw (A2.south east) -- (A2.north west);

  \node[left = 1.5em of A1] (x) {\scriptsize{$\code{x} = \code{0x01}$}};
  \draw[->] (x) --(A1);
  \node (x0) [above = 0pt of x]{};

  \node[part] (B1) [below = 2.5em of A1] {};
  \node[part] (B2) [right = 0pt of B1, label=right:\scriptsize{$\code{0x02}$}] {};

  \node[part] (C1) [below = 2.5em of B1] {};
  \node[part] (C2) [right= 0pt of C1, label=right:\scriptsize{$\code{0x03}$}] {};

  \draw (C1.south west) -- (C1.north east);
  \draw (C1.south east) -- (C1.north west);

  \node[part] (D1) [below = 2.5em of C1] {};
  \node[part] (D2) [right= 0pt of D1, label=right:\scriptsize{$\code{0x04}$}] {};
  \node[left = 1.5em of D1] (y) {\scriptsize{$\code{y} = \code{0x04}$}};
  \draw[->] (y) --(D1);

  \draw (D2.south west) -- (D2.north east);
  \draw (D2.south east) -- (D2.north west);

  \node[part] (E1) [below = 2.5em of D1] {};
  \node[part] (E2) [right= 0pt of E1, label=right:\scriptsize{$\code{0x05}$}] {};
  \node (E0) [below=0pt of E2]{};

  \draw (E1.south west) -- (E1.north east);
  \draw (E1.south east) -- (E1.north west);

  \draw[->] (A1.center) -- (B1);
  \node[rotate=270,anchor=north,yshift=2pt] (n1) at ($(A1)!0.5!(B1)$) {\scriptsize{\code{next}}};

  \draw[->] (B1.center) -- (C1);
  \node[rotate=270,anchor=north,yshift=2pt] (n2) at ($(B1)!0.5!(C1)$) {\scriptsize{\code{next}}};

  \draw[->] (D1.center) -- (E1);
  \node[rotate=270,anchor=north,yshift=2pt] (n4) at ($(D1)!0.5!(E1)$) {\scriptsize{\code{next}}};

  \draw[<-] (A2) -- (B2.center);
  \node[rotate=90,anchor=north,yshift=2pt] (p1) at ($(A2)!0.5!(B2)$) {\scriptsize{\code{prev}}};

  \draw[<-] (B2) -- (C2.center);
  \node[rotate=90,anchor=north,yshift=2pt] (p2) at ($(B2)!0.5!(C2)$) {\scriptsize{\code{prev}}};

  \draw[<-] (D2) -- (E2.center);
  \node[rotate=90,anchor=north,yshift=2pt] (p4) at ($(D2)!0.5!(E2)$) {\scriptsize{\code{prev}}};

  \node[draw,inner sep=2pt,label=above:{\em stack},fit={(x0) (y) (x0.north|-E0.south)}] {};
  \node[draw,inner sep=2pt,label=above:{\em heap}, fit=(A0) (E0) (a1) (n1)] {};

\end{tikzpicture}
}

\newcommand{\invtraces}[1]{
\begin{tikzpicture}[scale=#1,transform shape]
    \tikzstyle{part} = [rectangle, draw=black!50, fill=black!20, inner sep=0.5pt, outer sep=0pt,minimum size=1em]

    \node[draw,label=above:\footnotesize{$s_1$}] (stk1) {
      \setlength\tabcolsep{1pt}
      \scriptsize
      \begin{tabular}{rll}
        \code{x} & = & \code{0x01} \\
        \code{tmp} & = & \code{0x02} \\
        \code{y} & = & \code{0x04} \\
        $\res$ & = & \code{0x01}
      \end{tabular}};
    \node[left=0.1em of stk1] (t1) {$t_{1}$};

    \node[draw,below=1.5em of stk1,label=above:\footnotesize{$s_2$}] (stk2) {
      \setlength\tabcolsep{1pt}
      \scriptsize
      \begin{tabular}{rll}
        \code{x} & = & \code{0x02} \\
        \code{tmp} & = & \code{0x03} \\
        \code{y} & = & \code{0x04} \\
        $\res$ & = & \code{0x02}
      \end{tabular}};
    \node[left=0.1em of stk2] (t2) {$t_{2}$};

    \node[draw,below=1.5em of stk2,label=above:\footnotesize{$s_3$}] (stk3) {
      \setlength\tabcolsep{1pt}
      \scriptsize
      \begin{tabular}{rll}
        \code{x} & = & \code{0x03} \\
        \code{tmp} & = & \code{0x04} \\
        \code{y} & = & \code{0x04} \\
        $\res$ & = & \code{0x03}
      \end{tabular}};
    \node[left=0.1em of stk3] (t3) {$t_{3}$};

    \node[part, anchor=north west] (A1) at ([xshift=1em] stk1.north east) {};
    \node[part] (A2) [right= 0pt of A1, label={[name=a1]right:\scriptsize{\code{0x01}}}] {};
    \node (A0) [above= 0pt of A1]{};

    \draw (A2.south west) -- (A2.north east);
    \draw (A2.south east) -- (A2.north west);

    \node[part] (B1) [below = 2.5em of A1] {};
    \node[part] (B2) [right = 0pt of B1, label=right:\scriptsize{\code{0x02}}] {};

    \node[part] (C1) [below = 2.5em of B1] {};
    \node[part] (C2) [right= 0pt of C1, label=right:\scriptsize{\code{0x03}}] {};

    \node[part] (D1) [below = 2.5em of C1] {};
    \node[part] (D2) [right= 0pt of D1, label=right:\scriptsize{\code{0x04}}] {};

    \node[part] (E1) [below = 2.5em of D1] {};
    \node[part] (E2) [right= 0pt of E1, label=right:\scriptsize{\code{0x05}}] {};
    \node (E0) [below=0pt of E2]{};

    \draw (E1.south west) -- (E1.north east);
    \draw (E1.south east) -- (E1.north west);

    \draw[->] (A1.center) -- (B1);
    \node[rotate=270,anchor=north,yshift=2pt] (n1) at ($(A1)!0.5!(B1)$) {\scriptsize{\code{next}}};

    \draw[->] (B1.center) -- (C1);
    \node[rotate=270,anchor=north,yshift=2pt] (n2) at ($(B1)!0.5!(C1)$) {\scriptsize{\code{next}}};

    \draw[->] (C1.center) -- (D1);
    \node[rotate=270,anchor=north,yshift=2pt] (n3) at ($(C1)!0.5!(D1)$) {\scriptsize{\code{next}}};

    \draw[->] (D1.center) -- (E1);
    \node[rotate=270,anchor=north,yshift=2pt] (n4) at ($(D1)!0.5!(E1)$) {\scriptsize{\code{next}}};

    \draw[<-] (A2) -- (B2.center);
    \node[rotate=90,anchor=north,yshift=2pt] (p1) at ($(A2)!0.5!(B2)$) {\scriptsize{\code{prev}}};

    \draw[<-] (B2) -- (C2.center);
    \node[rotate=90,anchor=north,yshift=2pt] (p2) at ($(B2)!0.5!(C2)$) {\scriptsize{\code{prev}}};

    \draw[<-] (C2) -- (D2.center);
    \node[rotate=90,anchor=north,yshift=2pt] (p3) at ($(C2)!0.5!(D2)$) {\scriptsize{\code{prev}}};

    \draw[<-] (D2) -- (E2.center);
    \node[rotate=90,anchor=north,yshift=2pt] (p4) at ($(D2)!0.5!(E2)$) {\scriptsize{\code{prev}}};

    \node[draw,inner sep=2pt,label=above:\scriptsize{$h_1,h_2,h_3$},fit=(A0) (E0) (a1) (n1)] {};

  \end{tikzpicture}
}

\newcommand{\subheapa}[1]{
  \begin{tikzpicture}[scale=#1,transform shape,draw, >=stealth]
    \tikzstyle{part} = [rectangle, draw=black!50, fill=black!20, inner sep=0.5pt, outer sep=0pt,minimum size=1em]
    \tikzstyle{npart} = [rectangle, inner sep=0.5pt, outer sep=0pt,minimum size=1em]

    \node[part] (A1) {};
    \node[part] (A2) [right = 0pt of A1, label={[name=a1] right:\scriptsize{\code{0x01}}}] {};

    \draw[red] (A2.south west) -- (A2.north east);
    \draw[red] (A2.south east) -- (A2.north west);

    \node[inner sep=2pt,text width=1em] (xl) [red, left = 1em of A1]
      {\baselineskip=0pt\scriptsize{$\code{x}$}, \scriptsize{$\res$}\par};
    \draw[->] (xl) -- (A1);

    \node[blue,draw,dashed,rounded corners=5pt,inner sep=3.5pt,fit=(A1) (A1) (a1) (a1),
          label=right:\scriptsize{$h'_1$}] (bnd) {};

    \node[part] (B1) [below = 2.5em of A1] {};
    \node[part] (B2) [right = 0pt of B1, label={[name=a2] right:\scriptsize{\code{0x02}}}] {};

    \node[red] (tmpl) [left = 1em of B1] {\scriptsize{\code{tmp}}};
    \draw[->] (tmpl) -- (B1);

    \node[part] (C1) [below = 2.5em of B1] {};
    \node[part] (C2) [right= 0pt of C1, label=right:\scriptsize{\code{0x03}}] {};

    \node[part] (D1) [below = 2.5em of C1] {};
    \node[part] (D2) [right= 0pt of D1, label={[name=a4] right:\scriptsize{\code{0x04}}}] {};

    \node (yl) [left = 1em of D1] {\scriptsize{\code{y}}};
    \draw[->] (yl) -- (D1);

    \node[part] (E1) [below = 2.5em of D1] {};
    \node[part] (E2) [right= 0pt of E1, label=right:\scriptsize{\code{0x05}}] {};

    \draw (E1.south west) -- (E1.north east);
    \draw (E1.south east) -- (E1.north west);

    \draw[->] (A1.center) -- (B1);
    \node[rotate=270,anchor=north,yshift=2pt] (n2) at ($(A1)!0.5!(B1)$) {\scriptsize{\code{next}}};

    \draw[->] (B1.center) -- (C1);
    \node[rotate=270,anchor=north,yshift=2pt] (n2) at ($(B1)!0.5!(C1)$) {\scriptsize{\code{next}}};

    \draw[->] (C1.center) -- (D1);
    \node[rotate=270,anchor=north,yshift=2pt] (n3) at ($(C1)!0.5!(D1)$) {\scriptsize{\code{next}}};

    \draw[->] (D1.center) -- (E1);
    \node[rotate=270,anchor=north,yshift=2pt] (n4) at ($(D1)!0.5!(E1)$) {\scriptsize{\code{next}}};

    \draw[<-] (A2) -- (B2.center);
    \node[rotate=90,anchor=north,yshift=2pt] (p1) at ($(A2)!0.5!(B2)$) {\scriptsize{\code{prev}}};

    \draw[<-] (B2) -- (C2.center);
    \node[rotate=90,anchor=north,yshift=2pt] (p2) at ($(B2)!0.5!(C2)$) {\scriptsize{\code{prev}}};

    \draw[<-] (C2) -- (D2.center);
    \node[rotate=90,anchor=north,yshift=2pt] (p2) at ($(C2)!0.5!(D2)$) {\scriptsize{\code{prev}}};

    \draw[<-] (D2) -- (E2.center);
    \node[rotate=90,anchor=north,yshift=2pt] (p2) at ($(D2)!0.5!(E2)$) {\scriptsize{\code{prev}}};

  \end{tikzpicture}
}

\newcommand{\subheapb}[1]{
  \begin{tikzpicture}[scale=#1,transform shape, draw, >=stealth]
    \tikzstyle{part} = [rectangle, draw=black!50, fill=black!20, inner sep=0.5pt, outer sep=0pt, minimum size=1em]
    \tikzstyle{npart} = [rectangle, inner sep=0.5pt, outer sep=0pt, minimum size=1em]

    \node[part] (A1) {};
    \node[part] (A2) [right= 0pt of A1, label={[name=a1] right:\scriptsize{\code{0x01}}}] {};

    \draw[red] (A2.south west) -- (A2.north east);
    \draw[red] (A2.south east) -- (A2.north west);

    \node[part] (B1) [below = 2.5em of A1] {};
    \node[part] (B2) [right = 0pt of B1, label={[name=a2] right:\scriptsize{\code{0x02}}}] {};

    \node[inner sep=2pt,text width=1em] (xl) [red, left = 1em of B1]
    {\baselineskip=0pt\scriptsize{$\code{x}$}, \scriptsize{$\res$}\par};
    \draw[->] (xl) -- (B1);

    \node[blue,draw,dashed,rounded corners=5pt,inner sep=3.5pt,fit=(A1) (B1) (a2) (a2),
          label=right:\scriptsize{$h'_2$}] (bnd) {};

    \node[part] (C1) [below = 2.5em of B1] {};
    \node[part] (C2) [right= 0pt of C1, label={[name=a3] right:\scriptsize{\code{0x03}}}] {};

    \node[red] (tmpl) [left = 1em of C1] {\scriptsize{\code{tmp}}};
    \draw[->] (tmpl) -- (C1);

    \node[part] (D1) [below = 2.5em of C1] {};
    \node[part] (D2) [right= 0pt of D1, label={[name=a4] right:\scriptsize{\code{0x04}}}] {};

    \node (yl) [left = 1em of D1] {\scriptsize{\code{y}}};
    \draw[->] (yl) -- (D1);

    \node[part] (E1) [below = 2.5em of D1] {};
    \node[part] (E2) [right= 0pt of E1, label=right:\scriptsize{\code{0x05}}] {};

    \draw (E1.south west) -- (E1.north east);
    \draw (E1.south east) -- (E1.north west);

    \draw[->] (A1.center) -- (B1);
    \node[rotate=270,anchor=north,yshift=2pt] (n2) at ($(A1)!0.5!(B1)$) {\scriptsize{\code{next}}};

    \draw[->] (B1.center) -- (C1);
    \node[rotate=270,anchor=north,yshift=2pt] (n2) at ($(B1)!0.5!(C1)$) {\scriptsize{\code{next}}};

    \draw[->] (C1.center) -- (D1);
    \node[rotate=270,anchor=north,yshift=2pt] (n3) at ($(C1)!0.5!(D1)$) {\scriptsize{\code{next}}};

    \draw[->] (D1.center) -- (E1);
    \node[rotate=270,anchor=north,yshift=2pt] (n4) at ($(D1)!0.5!(E1)$) {\scriptsize{\code{next}}};

    \draw[<-] (A2) -- (B2.center);
    \node[rotate=90,anchor=north,yshift=2pt] (p1) at ($(A2)!0.5!(B2)$) {\scriptsize{\code{prev}}};

    \draw[<-] (B2) -- (C2.center);
    \node[rotate=90,anchor=north,yshift=2pt] (p2) at ($(B2)!0.5!(C2)$) {\scriptsize{\code{prev}}};

    \draw[<-] (C2) -- (D2.center);
    \node[rotate=90,anchor=north,yshift=2pt] (p2) at ($(C2)!0.5!(D2)$) {\scriptsize{\code{prev}}};

    \draw[<-] (D2) -- (E2.center);
    \node[rotate=90,anchor=north,yshift=2pt] (p2) at ($(D2)!0.5!(E2)$) {\scriptsize{\code{prev}}};

  \end{tikzpicture}
}

\newcommand{\subheapc}[1]{
  \begin{tikzpicture}[scale=#1,transform shape, draw, >=stealth]
    \tikzstyle{part} = [rectangle, draw=black!50, fill=black!20, inner sep=0.5pt, outer sep=0pt,minimum size=1em]
    \tikzstyle{npart} = [rectangle, inner sep=0.5pt, outer sep=0pt,minimum size=1em]

    \node[part] (A1) {};
    \node[part] (A2) [right= 0pt of A1, label={[name=a1] right:\scriptsize{\code{0x01}}}] {};

    \draw[red] (A2.south west) -- (A2.north east);
    \draw[red] (A2.south east) -- (A2.north west);

    \node[part] (B1) [below = 2.5em of A1] {};
    \node[part] (B2) [right = 0pt of B1, label=right:\scriptsize{\code{0x02}}] {};

    \node[part] (C1) [below = 2.5em of B1] {};
    \node[part] (C2) [right= 0pt of C1, label={[name=a3] right:\scriptsize{\code{0x03}}}] {};

    \node[inner sep=2pt,text width=1em] (xl) [red, left = 1em of C1]
    {\baselineskip=0pt\scriptsize{$\code{x}$}, \scriptsize{$\res$}\par};
    \draw[->] (xl) -- (C1);

    \node[blue,draw,dashed,rounded corners=5pt,inner sep=3.5pt,fit=(A1) (C1) (a3) (a3),
          label=right:\scriptsize{$h'_3$}] {};

    \node[part] (D1) [below = 2.5em of C1] {};
    \node[part] (D2) [right= 0pt of D1, label={[name=a4] right:\scriptsize{\code{0x04}}}] {};

    \node[red,text width=1em,inner sep=2pt] (yl) [left = 1em of D1]
      {\baselineskip=0pt\scriptsize{$\code{tmp}$}, \scriptsize{$\code{y}$}\par};
    \draw[->] (yl) -- (D1);

    \node[part] (E1) [below = 2.5em of D1] {};
    \node[part] (E2) [right= 0pt of E1, label=right:\scriptsize{\code{0x05}}] {};

    \draw (E1.south west) -- (E1.north east);
    \draw (E1.south east) -- (E1.north west);

    \draw[->] (A1.center) -- (B1);
    \node[rotate=270,anchor=north,yshift=2pt] (n2) at ($(A1)!0.5!(B1)$) {\scriptsize{\code{next}}};

    \draw[->] (B1.center) -- (C1);
    \node[rotate=270,anchor=north,yshift=2pt] (n2) at ($(B1)!0.5!(C1)$) {\scriptsize{\code{next}}};

    \draw[->] (C1.center) -- (D1);
    \node[rotate=270,anchor=north,yshift=2pt] (n3) at ($(C1)!0.5!(D1)$) {\scriptsize{\code{next}}};

    \draw[->] (D1.center) -- (E1);
    \node[rotate=270,anchor=north,yshift=2pt] (n4) at ($(D1)!0.5!(E1)$) {\scriptsize{\code{next}}};

    \draw[<-] (A2) -- (B2.center);
    \node[rotate=90,anchor=north,yshift=2pt] (p1) at ($(A2)!0.5!(B2)$) {\scriptsize{\code{prev}}};

    \draw[<-] (B2) -- (C2.center);
    \node[rotate=90,anchor=north,yshift=2pt] (p2) at ($(B2)!0.5!(C2)$) {\scriptsize{\code{prev}}};

    \draw[<-] (C2) -- (D2.center);
    \node[rotate=90,anchor=north,yshift=2pt] (p2) at ($(C2)!0.5!(D2)$) {\scriptsize{\code{prev}}};

    \draw[<-] (D2) -- (E2.center);
    \node[rotate=90,anchor=north,yshift=2pt] (p2) at ($(D2)!0.5!(E2)$) {\scriptsize{\code{prev}}};
  \end{tikzpicture}
}

\section{Illustration}
\label{sec:motiv}

\begin{figure}[t]
  \centering
  \begin{lstlisting}[emph={L1,L2,L3},xleftmargin=1.cm,escapechar=|]
typedef struct Node {
  struct node *next, *prev;
} Node;

Node *concat(Node *x, Node *y) {
  [L1] |\label{line:cc_pre}|
  if (x == NULL) {
    [L2] |\label{line:cc_post_1}|
    return y;
  } else {
    Node *tmp = concat(x->next, y);|\label{line:rec_call}|
    x->next = tmp;
    if (tmp) tmp->prev = x;
    [L3] |\label{line:cc_post_2}|
    return x;
  }
}
\end{lstlisting}
\caption{Concatenating two doubly linked lists}
\label{fig:concat}
\end{figure}

We describe {\tool} using the function \code{concat} shown in Figure~\ref{fig:concat}.
This function recursively concatenates two doubly linked lists \code{x} and \code{y} and returns (i) \code{y} when \code{x} is empty or (ii) a new doubly linked list by appending \code{y} to the tail of \code{x}.

Although simple, \code{concat} requires several subtle preconditions over its inputs to work properly.
First, \code{x} must be a $\nil$-terminated list, i.e., the \code{next} field of its tail node is \code{NULL}, otherwise \code{concat} may not terminate when \code{x} contains a cycle or may refer to an unallocated memory region when the \code{next} field of \code{x}'s tail node is a dangling pointer.
Second, \code{x} and \code{y} must be non-overlapping, i.e., point to lists in separate memory regions, otherwise the resulting list contains a cycle.
These conditions can be difficult to analyze or even to specify because they involve dynamically-allocated data structures and their separations in memory.
{\tool} aims to automatically discover such preconditions at program entrances and, more generally, invariant properties at arbitrary program locations, including postconditions and loop invariants.

\subsection{Heap Predicates}
\label{sec:hpredicates}

{\tool} infers invariants expressed as formulae in separation logic (SL) to describe properties of heap-manipulating programs.
Comparing to existing works for heap programs~\cite{jones:popl82,sagiv:toplas2002:parametric}, SL
provides concise and expressive syntax
and semantics to describe memory (shape) information~\cite{reynolds:lics02:sl,ohearn:csl01:sl}.

To analyze heap programs, SL works often use \emph{inductive heap predicates} to compactly represent recursively-defined data structures.
For \code{concat},
we use the predicate $\Dll$ to define doubly linked lists:

\begin{tabular}{lcl}
$\formPred{\Dll}{hd,pr,tl,nx}$&$\definedas$ & $(\emp \wedge hd{=}nx \wedge pr{=}tl)$\\
&$\vee$& $(\exists u.\, \formPtr{}{hd}{u,pr} \sep \formPred{\Dll}{u,hd,tl,nx})$
\end{tabular}

\noindent The parameters $hd, tl, pr$, and $nx$ point to the list's head, tail, previous, and next element, respectively.
The definition of $\Dll$ uses the built-in predicate  $\emp$ to represent an empty heap, e.g., a \code{NULL} list, and  the singleton predicate $\formPtr{\code{Node}}{x}{nx,pr}$ to denote a memory cell that a variable $x$ of type $\code{Node}$ shown in Figure~\ref{fig:concat} points to ($nx$ and $pr$ correspond to the $\code{next}$ and $\code{prev}$ fields of $x$, respectively\footnote{When the context is clear, we simply use $\formPtr{}{x}{nx,pr}$ for $\formPtr{\code{Node}}{x}{nx,pr}$.}).

Conceptually, $\Dll$ states that a
doubly linked list is either an empty or a non-empty list, which is recursively defined
by having the head $hd$ point to a doubly linked list whose head node is $u$.
In the latter case, the \emph{separating conjunction connector}
$\sep$ specifies the separation of memory regions modeled by $hd$'s singleton
predicate and $u$'s $\Dll$ predicate, i.e., the heaplets of $hd$ and $u$ are disjoint.

{\tool} uses such heap predicates to discover invariants and specifications of heap programs.
For \code{concat}, {\tool} uses $\Dll$ to generate the precondition on line~\ref{line:cc_pre} and the postcondition on line~\ref{line:cc_post_1} and ~\ref{line:cc_post_2} in Figure~\ref{fig:concat} as
\[
  \begin{aligned}
    \code{pre}  =~& \exists p, u, v.\, \formPred{\Dll}{\code{x},p,u,\nil} \sep
    \formPred{\Dll}{\code{y},\nil,v,\nil}\\
    \code{post} =~& \exists v.\, \formPred{\Dll}{\code{y},\nil,v,\nil}
    \wedge \code{x}{=}\nil \wedge \res{=}\code{y} ~\vee\\
    & \exists p,u,v.\,
    \formPred{\Dll}{\code{x},p,u,\code{y}} \sep
    \formPred{\Dll}{\code{y},u,v,\nil} \wedge \res{=}\code{x}
  \end{aligned}
\]

These pre and postconditions form a valid specification for \code{concat}.
The precondition requires that inputs $\code{x}$ and $\code{y}$ point to two disjoint, $\nil$-terminated doubly linked lists.
Note that unlike $\code{y}$, the $\Dll$ of $\code{x}$ shows that it can take any arbitrary
previous pointer, i.e., the existential argument $p$, because this pointer changes across the recursive call on line~\ref{line:rec_call}.

The postcondition ensures two exit conditions: (i) when $\code{x}$ is empty, the return value $\res$ is $\code{y}$, and (ii) otherwise, $\res = \code{x}$ is the result of appending $\code{y}$ to $\code{x}$ by changing the next element of $\code{x}$'s predicate from $\nil$ to $\code{y}$. Also, note that the previous field of $\code{y}$ now points to the tail element of $\code{x}$. Lastly, the postcondition states that the separation of the heaps of  $\code{x}$ and $\code{y}$ is preserved, i.e., $\code{concat}$ only changes the field values of the lists and does not alter the allocated memory.

Inductive heap predicates such as $\Dll$ are standard in SL (e.g., provided by the users or predefined in an analyzer~\cite{jacobs:nfm11:verifast,magill:cav08:thor,brotherston:popl16,le:cav14:s2}) and compactly capture crucial shape properties (e.g., doubly linked lists are acyclic).
In addition, compared to normal, non-SL predicates such as \code{isOdd} or $x \ge y$, checking SL heap predicates is nontrivial because we have to ``unfold'' data structures recursively and find concrete values to instantiate the existential quantifiers, e.g., the pre and postconditions above require finding the correct quantified variables and parameters in $\Dll$.

\subsection{Traces}

\begin{figure}[t]
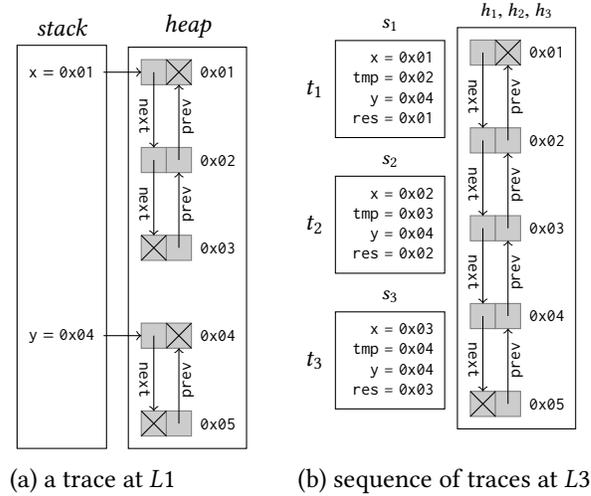

  \begin{tabular}{ll}
    \setlength\tabcolsep{0pt}
    \begin{minipage}{0.40\linewidth}
      \pretraces{0.95}
    \end{minipage}
    &
    \begin{minipage}{0.40\linewidth}
      \invtraces{0.95}
    \end{minipage}
    \\
    (a) a trace at $L1$
    & (b) sequence of traces at $L3$
  \end{tabular}
  \caption{Traces collected at $L1$ and $L3$ in $\code{concat}$.}
  \label{fig:traces}
 \end{figure}

Given a program annotated with locations of interest, {\tool} runs the program on sample inputs to collect execution traces.
Currently, we use the LLDB debugger~\cite{web:lldb} to observe execution traces containing memory addresses and values of the variables in scope at considered locations\footnote{These "traces", which are snapshots of program states, are often referred to as the \emph{stack-heap models} in SL literature, which we review in Section~\ref{sec:sp}.}.

Figure~\ref{fig:traces}a shows two doubly linked lists \code{x} and \code{y} of size 3 and 2, respectively.
When running on these inputs, {\tool} records traces such as those given in Figures~\ref{fig:traces}.
Figure~\ref{fig:traces}a shows the trace obtained in the first iteration of \code{concat} at $L1$.
The trace contains information about both the {\em stack}, containing values of variables accessible at this location, e.g., $\code{x=0x01}, \code{y=0x04}$, and the {\em heap}, containing allocated memory
cells reachable from the stack's variables, e.g., $\code{0x01} {\mapsto} \code{Node\{next:0x02;prev:nil\}}$.

Figure~\ref{fig:traces}b shows three set of traces collected at $L3$ for the first three iterations of \code{concat}.
Note that the values of \code{x} and \code{tmp} are different in the stacks because
\code{x} and \code{tmp} change across the recursive calls in $\code{concat}$.
However, the heap is similar because \code{concat} does not change the heap, e.g., delete or create cells, and all
memory cells are still reachable from the stack variables.
Moreover, the stack at $L3$ contains a \emph{ghost} variable $\res$, which
stores the return value of the function\footnote{This value is captured when the LLDB
debugger steps out of the function and jumps back to its call site.}.

\subsection{Inference}
\label{sec:motiv:infer}
\begin{figure}[t]
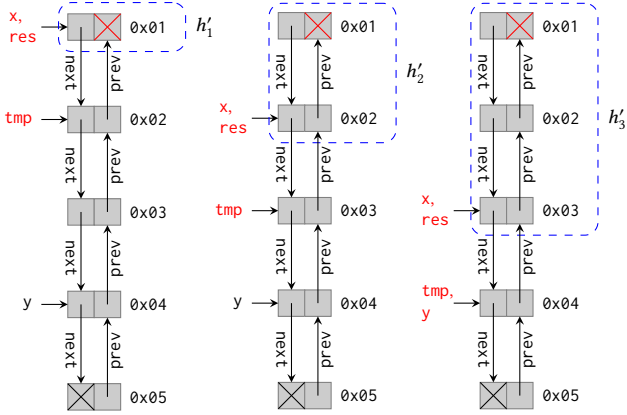

  \begin{minipage}{0.32\linewidth}
    \subheapa{1}
  \end{minipage}
  \begin{minipage}{0.32\linewidth}
    \subheapb{1}
  \end{minipage}
  \begin{minipage}{0.32\linewidth}
    \subheapc{1}
  \end{minipage}
  \caption{Sub-heaps of $\code{x}$ (in blue dashed boxes) and their boundaries
       ($\nil$ and the variables in red). The common boundary of these sub-heaps is
       $\{ \code{x}, \code{tmp}, \res, \nil \}$.}
  \label{fig:subheaps}
\end{figure}

{\tool} infers invariants consisting of SL predicates such as $\Dll$ over variables at a location of interest.
For each (pointer) variable, {\tool} explores relevant memory regions in observed traces to compute invariants for that variable.
Finally, {\tool} combines the computed invariants to model the whole explored memory regions.

\paragraph{Postcondition} We now show how {\tool} computes the postcondition at $L3$ using the
predicate $\Dll$ and the traces shown in Figure~\ref{fig:traces}b.
For demonstration we assume that {\tool} analyzes the variables at $L3$ in the order $\code{x}, \code{tmp}, \code{y}, \res$.

From given traces, {\tool} first computes the sub-heaps of \code{x} and their boundaries.
The \emph{sub-heaps} contain memory cells reachable from \code{x} but not pointed to by other stack variables.
The \emph{boundaries} of \code{x}'s sub-heaps contain \code{x} itself, the
$\nil$ pointer if it is reachable from \code{x}, and all variables reachable from \code{x} or its aliasing pointers.
Next, {\tool} takes the intersection of the boundaries to obtain the \emph{common boundary}, which consists of variables used to compute invariants in the next step.
Figure~\ref{fig:subheaps} shows the computed sub-heaps $h_1' \,{=}\, \{ \code{0x01} \}$, $h_2' \,{=}\, \{
\code{0x01}, \code{0x02} \}$, and $h_3' \,{=}\, \{ \code{0x01}, \code{0x02},
\code{0x03} \}$ over the three traces of \code{x} and their respective
boundaries $\{\code{x}, \res, \nil, \code{tmp}\}$, $\{\code{x}, \res, \nil, \code{tmp}\}$,
and $\{\code{x}, \res, \nil, \code{tmp}, \code{y}\}$.
Their common boundary is $\{\code{x}, \res, \nil, \code{tmp}\}$.

From computed sub-heaps and boundary variables, {\tool} searches the
predefined predicates for formulae that are consistent with sub-heap traces using boundary variables.
For each predicate, {\tool} creates \emph{candidate} formulae by instantiating predicate parameters with boundary variables.
It does so by enumerating different subsets of boundary variables as predicate parameters.
For subsets of size smaller than the number of parameters, {\tool} introduces fresh existential variables to instantiate the predicate.
In our example, {\tool} enumerates formulae such as $\exists u_1.\, \formPred{\Dll}{\code{x},\nil,\code{tmp},u_1}, \exists u_1.\, \formPred{\Dll}{\code{x},\nil,u_1,\code{tmp}}, \dots$.

Next, {\tool} then uses an  SMT-based model checker to check each candidate formula
against the given sub-heaps. The checker either refutes the formula, which is
then discarded, or accepts it, which {\tool} then considers as a valid formula
over the sub-heaps. Intuitively, accepted formulae represent partial invariants
computed from memory regions, e.g., the sub-heaps, related to the analyzed
variable.

In our example, among the generated candidates, the checker accepts the formula $F_x = \exists u_1, u_2.\,
\formPred{\Dll}{\code{x},u_1,u_2,\code{tmp}}$.
This formula shows that $\code{x}$ is a doubly linked list to the {\em next} pointer $\code{tmp}$.
The existential variables $u_1, u_2$ indicate that {\tool} cannot find concrete
stack or $\nil$ variables for the second and third parameters of $\Dll$ from the traces.

Although $F_x$  holds over the given sub-heaps, it might not generalize to the entire heap in the observed traces.
Thus, when analyzing $F_x$, the checker also computes a \emph{residual heap}, which represents the part of the heap that is not modeled by $F_x$.
The checker also computes a mapping from existential variables to concrete memory addresses from given traces.
{\tool} propagates these details to improve the analyses of other variables in subsequent iterations.

{\tool} now continues with the other variables $\code{tmp}, \code{y}, \res$ using the described steps and computed information (residual heaps and address mappings).
For $\code{tmp}$, {\tool} computes the sub-heaps and boundary $\{ \code{tmp},
\code{x}, \res, \code{y} \}$, and obtains then the formula $F_{\code{tmp}} = \exists
u_3.\, \formPred{\Dll}{\code{tmp}, \code{x}, u_3, \code{y}}$,
which indicates a doubly linked list from $\code{tmp}$ to the next pointer $\code{y}$.
Also, observe that the previous pointer points to $\code{x}$, showing the connection between this list and the one modeled by $F_x$.

Similarly, for the last two variables $\code{y}$ and $\code{res}$, {\tool} obtains the formulae $F_y =\exists u_4, u_5.\, \formPred{\Dll}{\code{y},u_4,u_5,\nil}$  and $F_{\code{res}} = \emp$.
$F_{\code{res}}$ is $\emp$ because every sub-heap reachable from $\res$ is empty in the traces observed in the last iteration.

The obtained formulae $F_{\code{x}}, F_{\code{y}}, F_{\code{tmp}}$, and $F_{\code{res}}$ model separate sub-heaps, thus {\tool} combines them using the $\sep$ operator to form a \emph{shape} invariant capturing the shape of the memory at $L3$, e.g., connections among separate heaplets:
\[
\begin{aligned}
  &\qquad \qquad \qquad F_{L3} = \exists u_1, u_2, u_3, u_4, u_5, \code{tmp}.\\
  &\formPred{\Dll}{\code{x},u_1,u_2,\code{tmp}} \sep
  \formPred{\Dll}{\code{tmp},\code{x},u_3,\code{y}} \sep
  \formPred{\Dll}{\code{y},u_4,u_5,\nil}.
\end{aligned}
\]
 Note that
 the constraint $F_{\code{res}}=\emp$ is discarded from the conjunction.
 Also note that the local variable $\code{tmp}$ is not in the scope of the function's exit, thus {\tool} considers it as an existential variable in $F_{L3}$.
In general, {\tool} only uses the function's parameters and the ghost variable $\res$ as free variables in the function's pre and postconditions.

{\tool} also examines analyzed information to find additional \emph{pure} (not related to memory) relations among the stack and existential variables in the inferred formula.
In this example, {\tool} determines that $\code{res}=\code{x}$, indicating that the return
value at $L3$ is $\code{x}$. It also infers aliasing information such as
$\code{x} \,{=}\, u_2$, $u_3 \,{=}\, u_4$ from the address mapping.
From these additional equalities, {\tool} derives the final result:
\[
  \begin{aligned}
    F'_{L3} = \exists u_1, u_3, u_5, \code{tmp}.\
    \formPred{\Dll}{\code{x},u_1,\code{x},\code{tmp}} ~\sep~ \\
    \formPred{\Dll}{\code{tmp},\code{x},u_3,\code{y}} ~\sep~
\formPred{\Dll}{\code{y},u_3,u_5,\nil} \wedge \res \,{=}\, \code{x}.
\end{aligned}
\]

\noindent This result $F'_{L3}$ is correct at $L3$ and even more precise than (stronger) the postcondition shown in Section~\ref{sec:hpredicates} when
$\code{x} \ne \nil$: $\exists p,u,v.\,
\formPred{\Dll}{\code{x},p,u,\code{y}} \sep
\formPred{\Dll}{\code{y},u,v,\nil} \wedge \res \,{=}\, \code{x}$.
The reason is because $\formPred{\Dll}{\code{x},u_1,\code{x},\code{tmp}} \sep
\formPred{\Dll}{\code{tmp},\code{x},u_3,\code{y}}$ in $F'_{L3}$
entails $\exists p,u.\, \formPred{\Dll}{\code{x},p,u,\code{y}}$
in the given postcondition.
The reversed direction of this entailment does not hold as it requires
a non-trivial condition $\code{x} \,{\neq}\, \code{y}$.

\paragraph{Precondition and Other Invariants} Using the same inference process over the traces obtained from the input \code{x}, \code{y} given
in Figure~\ref{fig:traces},  {\tool} infers the precondition at location $L1$
and the invariant at location $L2$ of \code{concat} as
\[
\begin{aligned}
  F'_{L1} = ~&\exists u_1, u_2, u_3, u_4.\\
  & \formPred{\Dll}{\code{x},u_1,u_2,\nil} \sep
  \formPred{\Dll}{\code{y},u_3,u_4,\nil} \wedge
  u_3 \,{=}\, \nil, ~\text{and}\\
  F'_{L2} = ~& \exists u_1, u_2.\,\\
  & \formPred{\Dll}{\code{y},u_1,u_2,\nil} \wedge
  u_1 \,{=}\, \nil \wedge \code{x} \,{=}\, \nil \wedge
  \res \,{=}\, \code{y}.
\end{aligned}
\]
These are the pre and postconditions shown in Section~\ref{sec:hpredicates}.
From these results, we obtain the specification
of \code{concat} because the complete post condition is the disjunction $F'_{L2}
\vee F'_{L3}$.
In general, {\tool} can compute invariants at arbitrary program locations by applying the described inference process to traces obtained at those locations.

Depending on the program, we could create scenarios where different orders of analyzed variables produce weaker results.
This is because the propagated residue information affects the computation of boundary variables and thus the instantiations of parameters in the predicates.
{\tool} prevents these scenarios by using a simple heuristic that only selects the next variables to analyze from those directly reachable from previously considered variables.
This gives a fixed order that works well in our experiments  (Section~\ref{sec:res}).

\def\semType{\mathsf{\code{Type}}}
\def\semVar{\mathsf{\code{Var}}}
\def\semVal{\mathsf{\code{Val}}}
\def\semLoc{\mathsf{\code{Loc}}}

\section{Separation Logic}
\label{sec:sp}

\begin{figure*}[t]
  \small
  \begin{tabular}{lll||lll}
    \textbf{Syntax}&&&\textbf{Semantics}&&\\
    \midrule
    $e\definedbnf$& $ k ~|~ x ~|~ -e ~|~ e_1 \,{+}\, e_2 ~|~ k \,{\cmul }\, e$&Integer exps&     $s,h \satisfies \emp$ & iff & $\fDom{h}\,{=}\,\setempty$ \\
    $a\definedbnf$ & $\nil ~|~ x$&Spatial exps & $s,h \satisfies \formPtr{\sort}{x}{t_1, ..., t_n}$ & iff & $\fDom{h}\,{=}\,\{ \evalVar{x}{s} \}$ and $h(\evalVar{x}{s}) = (\sort, \{\evalVar{t_1}{s}, ...,\evalVar{t_n}{s}\})$\\
    $\Pi \definedbnf$ & $a_1 \,{=}\, a_2 ~|~ e_1 \,{=}\, e_2 ~|~ e_1 \,{<}\, e_2 ~|~$&Pure formulae & $s,h \satisfies \formPred{\P}{t_1,...,t_n}$ & iff & $s,h \satisfies F$, where $F \DefRightarrow \formPred{\P}{t_1,...,t_n}$ \\
       &$ \neg\Pi ~|~ \Pi_1 \,{\wedge}\, \Pi_2 ~|~ \exists x.\, \Pi$&&$s,h \satisfies \Sigma_1 * \Sigma_2$ & iff & $\exists h_1, h_2.\, h_1 \,{\hdisjoins}\, h_2 \wedge h_1 \,{\hunions}\, h_2\,{=}\,h \wedge s,h_1 \,{\satisfies}\, \Sigma_1 \wedge  s,h_2 \,{\satisfies}\, \Sigma_2$ \\
    $\Sigma \definedbnf$ & $\emp ~|~ \formPtr{\sort}{x}{t_1,...,t_n}  ~|~ $&Spatial formulae & $s,h \satisfies \Sigma \wedge \Pi$ & iff & $\evalForm{\Pi}{s}\,{=}\,\true$ and $s,h \satisfies \Sigma$\\
    &$\formPred{\P}{t_1,...,t_n} ~|~  \Sigma_1 \sep \Sigma_2$&& $s,h \satisfies \exists x.\,F$ & iff & $\exists v \,{\in}\, \Val.\, \modelExt{s}{x}{v},h \satisfies F$\\
    $F \definedbnf$ & $\Sigma ~|~ \Pi ~|~ \Sigma \wedge \Pi ~|~ \exists x.\,F$&SL formulae
  \end{tabular}
  \caption{Syntax and semantics of symbolic-heap SL formulae.}
  \label{fig:SLsyntax}
\end{figure*}

SL~\cite{jones:popl82,sagiv:toplas2002:parametric} has been actively used to reason about imperative programs that manipulate data structures.
Crucially, SL uses the separating conjunction operation $\sep$ to describe the separation of computer memory, i.e., the assertion $p \sep q$ states that $p$ and $q$ hold for \emph{disjoint} memory regions.
Moreover, SL is often equipped with the ability for users to define inductive heap predicates (such as the predicate $\Dll$ used in Section~\ref{sec:motiv} for doubly linked list).
The combination of the $\sep$ operator and heap predicates make SL expressive enough to model various types of data structures.

Figure~\ref{fig:SLsyntax} shows the syntax and semantics of the SL formulae we consider in this work.
These represent the standard symbolic-heap fragment of SL~\cite{brotherston:popl16,ta:fm16,ta:popl18} with user-defined inductive heap predicates.

\paragraph{Syntax} We denote $x$ as a variable, $k,e$ as an integer constant and an integer expression, respectively, $\nil$ as a constant denoting a dangling memory address (\emph{null}), and $a$ as a spatial expression modeling a memory address.
The predicate $\emp$ models an empty heap,
the singleton heap predicate  $\formPtr{\sort}{x}{t_1,...,t_n}$ models an $n$-field data structure type $\sort$ where $x$ points to,
and the inductive heap predicate $\formPred{\P}{t_1,...,t_n}$ models a recursively defined data structure.
The \emph{spatial} formulae $\Sigma$ consist of these predicates and their compositions using the  $\sep$ separating conjunction operator.
$\Pi$ denotes \emph{pure formulae} in linear arithmetic, which do not contain any predicates.
Note that we can negate the presented formulae to obtain formulae involving disjunctions, universal quantifiers, and other comparison relations.

\paragraph{Semantics} Given a set $\semVar$ of variables, $\semType$ of types, $\semVal$
of values, and $\semLoc \,{\subset}\, \semVal$ of memory addresses, an SL
\emph{stack-heap model}, i.e., concrete trace, is a pair of
a \emph{stack model} $s$, which is a function $s{:} ~ \semVar \rightarrow \semVal$, and a \emph{heap model} $h$, which is a partial function $h{:} ~ \semLoc \rightharpoonup (\semType \times \Val^*)$.
We write $\evalForm{\Pi}{s}$ to denote the valuation of a formula $\Pi$ under the stack model $s$ and $\model{s, h}{F}$ to denote a model $s,h$ satisfies a formula $F$.
Moreover, $\fDom{h}$ denotes the domain of $h$,  $h_1 \,{\hdisjoins}\, h_2$ denotes $h_1$ and $h_2$ have different domains, and $ h_1 \,{\hunions}\, h_2$ denotes the union of two disjoint heaps $h_1$ and $h_2$, and $\modelExt{f}{x}{y}$ indicates a function like $f$ but returns $y$ for input $x$.
We also define the heap union and difference operators over two sequences of stack-heap models as $(s_i, h_i)_{i=1}^n \oplus (s_i, h'_i)_{i=1}^n \definedas (s_i, h_i \hunions h'_i)_{i=1}^n$ and $(s_i, h_i)_{i=1}^n \setminus (s_i, h'_i)_{i=1}^n \definedas (s_i, h_i \setminus h'_i)_{i=1}^n$, respectively.

\paragraph{Model Checking}  We follow the technique given in \cite{brotherston:popl16} to implement a model checker, which checks if a formula $F$ is satisfied by a stack-heap model $s, h$ and returns a residual heap $h'$, i.e., memory regions in  $h$ not modeled by $F$, and an instantiation $\inst$ that maps existential variables in $F$ to concrete addresses in the model.
These checking and instantiation tasks are encoded as logical formulae solvable  using the Z3 SMT solver~\cite{demoura:tacas08:z3}.

Note that the model checking technique proposed in ~\cite{brotherston:popl16} does not return the instantiation $\inst$, which is needed by {\tool} to compute equalities among variables in $F$.
To obtain $\inst$, we slightly redefine the problem with a new satisfaction relation:

\begin{definition}[Satisfaction Relation with Existential Instantiation]
  \label{def:sat_inst}
  The relation $\model[\inst]{s, h}{F}$ is the satisfaction relation
  $\model{s, h}{F}$ except that the value of an existential variable in $F$
  is obtained from the instantiation $\inst$, which is a function from $\semVar$
  to $\semVal$ similar to the stack model.
\end{definition}

We also lift this relation to sequences of stack-heap models and instantiations as $\model[(\inst_i)_{i=1}^n]{(s_i, h_i)_{i=1}^n}{F} \definedas \forall i.\; \model[\inst_i]{s_i, h_i}{F}$.

\begin{definition}[Symbolic-heap Model Checking]\label{def:model_check}
  A reduction $\modelcheck{s, h}{F}{h', \inst}$ is valid if $h' \subseteq h$
  and $\model[\inst]{s, h \setminus h'}{F}$.
\end{definition}

\noindent Definition \ref{def:model_check} redefines the model checking reduction
relation to return, in addition to the residual heap model $h'$, an instantiation $\inst$ of existential variables in $F$ that satisfies the relation $\model[\inst]{s, h \setminus h'}{F}$ in Definition \ref{def:sat_inst}.

\section{The {\tool} Algorithm}
\label{sec:alg}

\def\Prog{\ensuremath{\mathsf{C}}}
\def\TestSuite{\ensuremath{\mathsf{T}}}
\def\sSH{\ensuremath{\mathsf{SH}}}
\def\sRem{\ensuremath{\mathsf{R}}}
\def\Pred{\ensuremath{\mathsf{P}}}
\def\sBoundary{\ensuremath{\mathsf{B}}}
\def\sArg{\ensuremath{\mathsf{A}}}
\def\ssArg{\ensuremath{\widehat{\sArg}}}
\def\sVar{\ensuremath{\mathsf{V}}}
\def\sInst{\ensuremath{\mathsf{I}}}
\newcommand{\powerset}{\ensuremath{\mathsf{PowerSet}}}
\newcommand{\perm}[2]{\ensuremath{\mathsf{Perm}_{#1}(#2)}}

\def\CollectTraces{\proc{CollectModels}}
\def\pGetVars{\proc{GetVars}}
\def\SplitHeap{\proc{SplitHeap}}
\def\InferAtom{\proc{InferAtom}}
\def\InferSingleton{\proc{InferSingleton}}
\def\InferPure{\proc{InferPure}}
\def\Validate{\proc{Validate}}
\def\pFresh{\proc{fresh}}
\def\pType{\proc{type}}

\begin{algorithm}[t]
  \small
  \caption{The main algorithm of {\tool}}
  \label{fig:Sling}
  \begin{algosummary}
    \Summary{Input:} A program {\Prog}, a set of predefined predicates {\Pred}, a test suite {\TestSuite}, and a program location $l$

    \Summary{Output:} A set of invariants at $l$
  \end{algosummary}

  \begin{algorithmic}[1]
    \State \Assign{\sSH}{\Call{\CollectTraces}{\Prog, l, \TestSuite}}
    \label{line:CollectTraces}
    \State \Assign{\sVar}{\Call{\pGetVars}{l}}
    \Comment{stack variables}
    \label{line:AnalysisBegin}
    \label{line:GetLocalVars}
    \State \Assign{\sRem}{\{(\emp, \sSH, \seq{\inst_i = \{\}}{i \in \TestSuite})\}}
    \Comment{initial result}
    \label{line:InferenceInit}
    \For{\Keyword{each ~ pointer} v \Keyword{in} \sVar}
    \label{line:RefineBegin}
    \State \Assign{\sRem'}{\{\}}
    \For{\Keyword{each} (F, \sSH, \sInst) \Keyword{in} \sRem}
    \State \Assign{\sSH_v, \sSH_r, \sBoundary}{\Call{\SplitHeap}{\sSH, v}}
    \Comment{$\sSH_v \oplus \sSH_r \equiv \sSH$}
    \label{line:SplitHeap}
    \State \Assign{\sRem_v}{\Call{\InferAtom}{v, \sSH_v, \sBoundary, \Pred}}
    \label{line:InferAtom}
    \For{\Keyword{each} (F_v, \sSH', \sInst') \Keyword{in} \sRem_v}
    \Comment{$\model[\sInst']{\sSH_v \setminus \sSH'}{F_v}$}
    \State \Assign{\sRem'}
    { \sRem' \cup \{ (F \sep F_v, \sSH_r \oplus \sSH', \sInst \oplus \sInst') \} }
    \label{line:ComputeNewRes}
    \EndFor
    \EndFor
    \State \Assign{\sRem}{\{ (\Call{\InferPure}{F, \sSH, \sInst}, \sSH, \sInst)
      ~|~ (F, \sSH, \sInst) \in \sRem' \}}
    \label{line:InferPure}
    \EndFor
    \label{line:RefineEnd}
    \label{line:AnalysisEnd}
    \State \Return{\sRem}
  \end{algorithmic}
\end{algorithm}

Algorithm~\ref{fig:Sling} shows the implementation of {\tool}.
Given a program {\Prog}, a set {\Pred} of predefined inductive heap predicates, a target location $l$ in {\Prog},  and a test suite {\TestSuite}, {\tool} returns a set $\sRem$ of SL formulae satisfied by observed traces at $l$.

{\tool} infers invariants using the three main phases described below.
In the following we use the term \emph{stack-heap models} to refer to concrete traces.

\paragraph{Model Collection (line~\ref{line:CollectTraces})}
{\tool} first calls {\CollectTraces} to collect all stack-heap models observed at location $l$ when running the program over the tests in {\TestSuite}.
{\CollectTraces} uses a software debugger such as LLDB to set a breakpoint at $l$ and inspect the memory layout when executing the program.
It then collects the set of stack-heap models $\sSH$ from the memory whenever hitting the breakpoint at $l$.

\paragraph{Inference (lines  \ref{line:AnalysisBegin}--\ref{line:AnalysisEnd})} After obtaining the stack-heap models at $l$, {\tool} performs a {\em heap} inference and then a {\em pure} inference to derive a set of results satisfied by the models.
{\tool} uses an iterative refinement process over the stack variables to infer invariants.
At each iteration, {\tool} updates the result set $\sRem$ with a tuple $(F, \sSH, \sInst)$,
where the formula $F$ holds for the models analyzed in the previous iteration, the set of stack-heap models $\sSH$ captures the residue of the initial heaps that are not modeled by $F$, and the sequence $\sInst$ contains existential instantiations which map the existential variables in $F$ to concrete memory addresses.

In each iteration, given a stack variable $v \in \sVar$ and a tuple $(F, \sSH, \sInst) \in \sRem$, {\tool} derives a set of atomic heap predicates (i.e., inductive heap predicates, singleton heap predicates, or $\emp$), which models the sub-heaps in $\sSH$ that contain memory cells reachable from $v$.
The heaps modeled by these predicates and $F$ are disjoint, thus we can strengthen $F$ with each predicate using the $\sep$ operator of SL.
Intuitively, {\tool} splits the original stack-heap models into multiple sub-heaps, which are pointed-to by distinct (non-aliasing) stack variables.
To model a sub-heap, {\tool} derives atomic formulae from the given predicates and the stack variables related to the sub-heap.
Sections~\ref{sec:split_heap} and~\ref{sec:infer_atom} describes the two sub-procedures {\SplitHeap} and {\InferAtom}, respectively.

In addition to finding invariants describing shape properties,  {\tool} infers equality constraints over stack variables in $\sVar$ to represent pure properties.
Section~\ref{sec:infer_pure} describes the {\InferPure} procedure that performs this step.

\paragraph{Validation} When we discover both pre and postconditions, we combine them to obtain program specifications.
We also leverage the \emph{frame rule} of SL to check if this combination is consistent with respect to the corresponding residual models.
Thus, when inferring invariants at multiple \code{return} statements, {\tool} has an additional step that combines and validates formulae inferred at these locations. We describe this step in Section \ref{sec:validate}.

\subsection{Heap Partitioning}
\label{sec:split_heap}

Given a sequence $\sSH$ of collected stack-heap models, {\tool} calls {\SplitHeap} to splits the heap in each model $s_i, h_i \in \sSH$ into smaller sub-heaps so that each of them can be modeled by atomic heap predicates.
Moreover, {\SplitHeap} returns the common boundary of these sub-heaps, which consist of the $\nil$ and stack variables that are subsequently used to determine the arguments for these atomic heap predicates.

{\SplitHeap} uses a depth-first search to traverse the pointer fields of memory cells from a $root$ pointer to partition the heap model $h_i$ into two non-overlapping parts: sub-heap $h'_i$ and the remaining sub-heap $h''_i = h_i \setminus h'_i$.
The sub-heap $h'_i$ contains memory cells reachable from the $root$ pointer variable up to the $\nil$ pointer or memory cells pointed to by other stack pointer variables.
We call these pointer variables or the $\nil$ pointer the {\em boundary} between the sub-heap $h'_i$ and the other memory regions in the heap $h_i$.
The sub-heap $h''_i$ may contain memory cells unreachable from $root$ and those reachable from $root$, but also pointed-to by other stack variables non-aliasing with $root$.

For the \code{concat} example, Figure~\ref{fig:subheaps} illustrates that the boundaries of the sub-heaps $h'_1$, $h'_2$, and $h'_3$ of the root
variable $\code{x}$ are  $\{\code{x}, \res, \nil, \code{tmp}\}$, $\{\code{x}, \res, \nil, \code{tmp}\}$, and $\{\code{x}, \res, \nil, \code{tmp}, \code{y}\}$, respectively.
Their common boundary is $\{\code{x}, \res, \nil, \code{tmp}\}$.

\subsection{Inferring Atomic Heap Predicates}
\label{sec:infer_atom}

\begin{algorithm}[t]
  \small
  \caption{{\InferAtom}: Inferring Atomic Predicates}
  \label{fig:InferAtom}
  \begin{algosummary}
    \Summary{Input:} A stack pointer $root$, its sub-models $\sSH_{root}$ and
    their common boundary $\sBoundary$, and a set of predefined predicates $\Pred$

    \Summary{Output:} A set of atomic formulae modeling the sub-models and their residue information
  \end{algosummary}

  \begin{algorithmic}[1]
    \def\indent{\hspace{1.2em}}
      \State \Assign{\sRem}{\{\}}
    \For{\Keyword{each} \formPred{\P}{t_1,...,t_n} \Keyword{in} \Pred}
    \Comment{Consider a predicate $\P$}
    \label{line:InferIndPredBegin}
    \State \Assign{\ssArg}{\{ \sArg ~|~ \sArg \in ~\powerset(\sBoundary)~
      \wedge |\sArg| \leq n
      \wedge root \in \sArg_i \}}
    \label{line:InferChooseSubset}
    \For{\Keyword{each} \sArg \Keyword{in} \ssArg}
    \Comment{Consider a subset of $\sBoundary$}
    \State \Assign{\{ u_1, ..., u_m \}_{m = n {-} |\sArg|}}{\Call{\pFresh}{n {-} |\sArg|}}
    \label{line:InferAtomMkFresh}
    \State \Assign{\sArg}{\sArg \cup \{ u_1, ..., u_m \}_{m = n {-} |\sArg|}}
    \label{line:InferAtomExtArgs}
    \For{\Keyword{each ~ permutation} (k_1,...,k_n) \Keyword{in} \perm{n}{\sArg}}
    \label{line:InferPermSet}
    \If{\forall 1 \,{\leq}\, i \,{\leq}\, n.\,
      k_i \in \sBoundary \rightarrow \pType(k_i) \subtype \pType(t_i)}
    \label{line:InferTypeCheck}
    \State \Assign{F}{\exists u_1, ..., u_m.\, \formPred{\P}{k_1,...,k_n}}
    \label{line:InferConstructF}
    \If{\forall s_i, h_i \in \sSH_{root}.\, \modelcheck{s_i, h_i}{F}{h'_i, \inst_i}}
    \label{line:InferCheckF}
    \State \Assign{\sRem}{\sRem \cup \{ (F, \seq{(s_i, h'_i)}{i}, \seq{\inst_i}{i}) \}}
    \label{line:InferUpdateRes}
    \EndIf
    \EndIf
    \EndFor
    \EndFor
    \EndFor
    \label{line:InferIndPredEnd}
    \If{\forall s_i, h_i \in \sSH_{root}.\, |h_i| = 1}
    \label{line:InferSingletonBegin}
    \State \Assign{\sRem}{\sRem \cup \Call{\InferSingleton}{root, \sSH_{root}}}
    \EndIf
    \label{line:InferSingletonEnd}
    \If{\sRem = \setempty}
    \label{line:InferEmp}
    \Assign{\sRem}{{\{(\emp, \sSH_{root}, \seq{\inst_i = \{\}}{i})\}}}
    \EndIf
    \State \Return{\sRem}
  \end{algorithmic}
\end{algorithm}

Given the sequence of sub-models $\sSH_{root}$ of the $root$ pointer and its boundary $\sBoundary$, the function $\InferAtom$ shown in Algorithm~\ref{fig:InferAtom} computes a set of atomic predicates satisfied by all sub-models in $\sSH_{root}$.
These atomic (shape) predicates consist of either (i) inductive heap predicates whose definitions are given in the set $\Pred$ (lines \ref{line:InferIndPredBegin}--\ref{line:InferIndPredEnd}), (ii) singleton predicates of the $root$ pointer when the heap size of all sub-models in $\sSH_{root}$ is 1 (lines \ref{line:InferSingletonBegin}--\ref{line:InferSingletonEnd}), or (iii) the $\emp$ predicate with $\sSH_{root}$ as the residual models when it cannot derive any predicates in the two former forms (line \ref{line:InferEmp}).

\paragraph{Inductive Heap Predicates} {\tool} discovers instances of each predefined predicate $\formPred{\P}{t_1,...,t_n} \in \Pred$.
For optimization, we filter the set $\Pred$ of predicate definitions to contain only those that have at least one parameter having the same type as the $root$ pointer.
Also, for simplicity of presentation, we assume that the parameters $t_1, ..., t_n$ of $\P$ are pointer types.

{\tool} chooses potential arguments of predicate $\P$ from the common boundary $\sBoundary$ of all sub-heaps in the sub-models $\sSH_{root}$.
It searches for these arguments from all permutation of $\sBoundary$'s subsets whose size is less than or equal to $n$ and contains $root$ (line \ref{line:InferChooseSubset}).
The inferred inductive predicates can contain as many stack variables as its arguments, thus we examine each subset in the ascending order of their size.
Also, to reduce the search space, we only consider a
permutation $(k_1, ..., k_n)$ if it is type-consistent with the parameters $t_1, ..., t_n$ of the predicate $\P$.
That is, if $k_i$ is a stack pointer variable in $\sBoundary$ then its type must be a subtype of the corresponding parameter $t_i$'s type (line \ref{line:InferTypeCheck}).

Next, we construct a formula $F$ from the inductive heap predicate $\formPred{\P}{k_1,...,k_n}$ (line \ref{line:InferConstructF}).
A formula $F$ is valid if it is successfully checked by all models in
$\sSH_{root}$ (line \ref{line:InferCheckF}). This validity check also returns a
residual heap $h'_i$ and an existential instantiation $\inst_i$ for each
stack-heap model $s_i, h_i$ in $\sSH_{root}$. They are respectively the member
of the sequence of residual models and the sequence of existential
instantiations associating with the valid formula $F$ as an inference result in
the set $\sRem$ (line \ref{line:InferUpdateRes}).

In the \code{concat} example, when selecting the argument set $\{ \code{x}, \res,
\code{tmp}, \nil \} \in \sArg$, the algorithm derives the formula $F_{x} =
\exists u_1.\, \formPred{\Dll}{u_1, \nil, \code{x}, \code{tmp}}$. This result
shows that $\code{x}$ is the last node of doubly linked lists whose head is
$u_1$ and its next pointer is $\code{tmp}$.
Moreover, $F_{x}$ models the whole sub-heaps of $\code{x}$ in Figure~\ref{fig:subheaps}, i.e., all residue models have empty heaps, when the existential variable $u_1$ is instantiated to the address $\code{0x01}$.
As another example, when considering another set of potential arguments $\{ \code{x}, \code{tmp} \} \in \sArg$, we infer $F_{x} = \exists u_1, u_2.\, \formPred{\Dll}{\code{x}, u_1, u_2, \code{tmp}}$, indicating that  $\code{x}$ is the head of a doubly linked list segment to $\code{tmp}$.

\paragraph{Singleton Heap Predicates} We only derive singleton heap
predicates of $root$ when there is a single memory cell in \emph{every} $root$'s
sub-model in $\sSH_{root}$ (line \ref{line:InferSingletonBegin}).
We consider a $\sort$-typed singleton predicate template of the form
$\formPtr{\sort}{root}{k_1, ..., k_n}$. The value of each field $k_i$
in the template is the common pointer variable (including $\nil$) pointing to the
corresponding field of every memory cell in $\sSH_{root}$. If there is no such
variable, we create a fresh existential variable for $k_i$ and update this variable's instantiation to the value of the corresponding field in each model.

\subsection{Pure Inference}
\label{sec:infer_pure}

The heap predicates derived in the previous steps mainly present the heap
memory, but not the relations of variables within a predicate and among
predicates in the overall results. In these results, the heap predicates are
solely related via the common stack variables in their arguments.

We infer additional pure constraints over arguments of the predicates
by searching for equality constraints over two different variables among
stack variables, existential variable, $\nil$, and the special variable $\res$
if we are inferring post-conditions which are satisfied by every stack model and existential instantiation.

For example, we use this inference to obtain the
relation $\res \,{=}\, \code{x}$ about the return value of $\code{concat}$ and other aliasing information, e.g., those shown in Section~\ref{sec:motiv:infer}.

\subsection{Validation}
\label{sec:validate}

When obtaining multiple postconditions (e.g., at each \code{return} statements), we combine them with the inferred preconditions to derive program specifications.
Next, we validate these specifications using frame rule of separation logic, i.e.,
\[
  \infer[]
  {\{ R \sep P \} ~\Prog~ \{ R \sep Q \}}
  {\{ P \} ~\Prog~ \{ Q \}}
\]
\noindent This rule says that if a triple ${\{ P \} \Prog \{ Q \}}$ is valid
(i.e., {\Prog} executes safely in the precondition $P$ and its post-states
satisfying the postcondition $Q$) then the triple ${\{ R \sep P \} \Prog \{ R
\sep Q \}}$, in which $R$ is a frame modeling memory regions that are not
manipulated by {\Prog}, also holds.

For example, if {\Prog} is a function, then $P$ and $Q$ are the pre and postconditions computed from the stack-heap models observed at the entry and exit of {\Prog}, respectively.
As another example, if {\Prog} is a loop body, then $P$ and $Q$, inferred from the models collected at the loop's head, are identical and considered as a loop invariant.

According to the frame rule, if the inferred conditions $P$ and $Q$ are valid, then we can expand the corresponding memory regions modeled by $P$ and $Q$ by the {\em same} memory regions non-overlapping with them to obtain the whole memories observed at the entry and exit {\Prog}. Otherwise, the inferred pair $P, Q$ is considered invalid (spurious).
Therefore, we can check that the residual models corresponding to $P$ and $Q$ are {\em unchanged} with respect to observed models to determine the validity of this result.

In \code{concat}, we obtain the precondition $F'_{L1}$ and two postconditions $F'_{L2}$ (when $\code{x} \,{=}\, \nil$) and $F'_{L3}$ (when  $\code{x} \,{\neq}\, \nil$).
For each pair of a model collected at $L1$ and
the corresponding model collected at $L2$ or $L3$, we check if the residual
heap in the model at $L1$, which is not captured by $F'_{L1}$, is the same as the
residual heap in model at $L2$ or $L3$, which is not captured by $F'_{L2}$ or $F'_{L3}$, respectively.
For example, given the model at $L1$ in Figure~\ref{fig:traces}a and its
corresponding model $t_1$ at $L3$ in Figure~\ref{fig:traces}b, the residual
heaps corresponding to $F'_{L1}$ and $F'_{L3}$ are both empty. On the other hand,
in the last iteration of \code{concat} (when $\code{x} \,{=}\, \nil$),
the residual heaps of $F'_{L1}$ and $F'_{L2}$ both contain three memory cells
\code{0x01}, \code{0x02}, and \code{0x03}.

\subsection{Complexity}
\label{sec:complexity}

{\tool} is exponential in the number of predicates and their parameters.
In addition, the complexity of the general heap model checking problem is \textsf{EXPTIME}~\cite{brotherston:popl16}.
Thus, checking predicates over combinations of variables over many collected stack-heap models can be slow.
To improve performance, {\tool} uses a type-checker (Algorithm~\ref{fig:InferAtom}, line~\ref{line:InferTypeCheck}) to eliminate variable combinations having inconsistent types to observed traces.
The experiments in Section~\ref{sec:eval} also show that only a few traces are needed to discover accurate invariants and the Z3-based model checker performs efficiently over these traces.

\section{Evaluation}
\label{sec:eval}

{\tool} is implemented in Python and uses the LLDB debugger~\cite{web:lldb} to collect traces at target program locations.
Below we evaluate {\tool} on C programs, but {\tool} also works with programs written in other languages supported by LLDB (e.g., C++ and Objective-C) or having debuggers capable of capturing memory information (e.g., JDB for Java~\cite{web:jdb}, PDB for
Python~\cite{web:pdb}, and GDB~\cite{web:gdb}).

Our experiments described below are conducted on a MacBook with 2.2GHz Intel CPU, 16 GB memory, and runs Mac OS.
The source code of {\tool} and experimental data are available at
\url{https://github.com/guolong-zheng/sling/}.

\subsection{Benchmark Programs}
\label{sec:benchmark}
We evaluate {\tool} using the VCDryad benchmark~\cite{web:vcdryad} consisting of 153 C heap-manipulating programs collected from various verification works, e.g., SV-COMP~\cite{beyer:tacas17:svcomp}, GRASShopper~\cite{piskac:tacas14:grasshopper} and AFWP~\cite{itzhaky:cav13}.
These programs range from those that manipulate standard data structures (e.g., heaps and trees) to functions from popular open source libraries (e.g., Glib, OpenBSD) and the Linux kernel that manipulate customized data structures.
Some of these programs have nontrivial bugs (e.g., causing segmentation faults) intended to test static analyzers.
We also use 4 programs\footnote{This benchmark has 6 programs, we use 4 of them and exclude the other 2 because they use concurrency which {\tool} currently does not support.} from~\cite{brotherston:popl16}.
These programs implement non-trivial algorithms using multiple data structures (e.g., the Schorr-Waite graph marking algorithm using binary trees).

In total, these benchmark programs contain a wide variety of structures including singly-linked lists, doubly-linked lists, sorted lists, circular lists, binary trees, AVL trees, red-black trees, heaps, queues, stacks, iterators, etc.
Moreover, these programs contain documented invariants (e.g., pre and postconditions such as those given in Section~\ref{sec:hpredicates}), which we use to evaluate {\tool}'s inferred invariants.

\begin{table*}
\centering
\caption{Experimental results. Programs denoted with $^*$ contain bugs preventing us to obtain traces.
  Programs denoted with $^{\dagger}$ cause {\tool} to timeout at certain locations.
  \emph{Italic} programs have locations that cannot be reached using random inputs.
  \textbf{Bold} programs contain locations with \code{free} statements that give invalid traces.
  SLL and DLL stand for Singly and Doubly Linked Lists, respectively. }
\label{tab:tab1}
\setlength\tabcolsep{2pt}
\small
\begin{tabular}{|l|c|c|c|c|c|c||c|c|c|}
\hline
\textbf{Programs}
& \multicolumn{6}{c||}{\textbf{Total}}
& \multicolumn{3}{c|}{\textbf{Avg. Per Inv}} \\
\cline{2-10} &
\textbf{LoC} & \textbf{iLocs} & \textbf{Traces} & \textbf{Invs} & \textbf{A/S/X} & \textbf{Time(s)} & \textbf{Single} & \textbf{Pred}  & \textbf{Pure}\\
 \hline

\footnotesize \begin{tabular}[c]{@{}l@{}} \textbf{SLL} (8):
  append, delAll, find, insert, reverse, insertFront, insertBack, copy
  \end{tabular}
  & 168 & 26 & 226 & 30 & 8/0/0 & 40.54 & 0.37 & 0.83 & 1.03 \\
  \hline

\footnotesize \begin{tabular}[c]{@{}l@{}}\textbf{Sorted List} (10):
  concat, find, findLast, insert, insertIter, delAll,\\
  reverseSort, insertionSort, mergeSort, quickSort$^*$
  \end{tabular}
  & 268 & 25 & 194 & 82 & 9/0/1 & 137.32 & 0.39 & 2.40 & 0.67 \\
  \hline

\footnotesize \begin{tabular}[c]{@{}l@{}}
                  \textbf{DLL} (12):
                append, concat, meld, delAll, insertBack, insertFront, midInsert, \\
                midDel, midDelError, midDelHd, midDelStar, midDelMid
                \end{tabular}
  & 160 & 31 & 168 & 238 & 12/0/0 & 399.12 & 0.46 & 1.68 & 3.93 \\
  \hline

\footnotesize \begin{tabular}[c]{@{}l@{}}
                  \textbf{Circular List} (4):
                  insertFront, insertBack, \textbf{\emph{delFront}}, \textbf{\emph{delBack}}
                \end{tabular}
  & 97 & 11 & 14 & 42(16) & 2/2/0 & 11.43 & 0.81       & 1.19     & 2.10 \\
  \hline

\footnotesize \begin{tabular}[c]{@{}l@{}}
                  \textbf{Binary Search Tree} (5):
                \emph{del}, findIter, \emph{find}, insert, rmRoot$^*$
                  \end{tabular}
  & 144 & 16     & 66     & 24    & 2/2/1  & 24.02
                          & 0.50       & 1.21     & 1.54 \\
  \hline

\footnotesize \begin{tabular}[c]{@{}l@{}}
                  \textbf{AVL Tree} (4):
                  \emph{avlBalance}, \emph{del}, findSmallest, insert
                \end{tabular}
  & 194 & 13     & 56       & 37      & 2/2/0 & 22.12
                          & 1.22       & 0.57     & 3.08 \\
  \hline

\footnotesize \begin{tabular}[c]{@{}l@{}}
                  \textbf{Priority Tree} (4):
                  \emph{del}, \emph{find}, \emph{insert}, \emph{rmRoot}
                \end{tabular}
      & 154 & 19     & 64       & 273     & 2/2/0 & 341.37
                          & 3.30       & 1.66     & 3.30    \\
  \hline

\footnotesize \begin{tabular}[c]{@{}l@{}}
                  \textbf{Red-black Tree} (2):
                  del$^*$, \emph{insert}
                  \end{tabular}
  & 287 & 11     & 70       & 63      & 0/1/1 & 44.8
                          & 2.10       & 1.08     & 8.11 \\
  \hline

\footnotesize \begin{tabular}[c]{@{}l@{}}
                \textbf{Tree Traversal} (5): traverseInorder, traversePostorder, traversePreorder, \\
                tree2list, tree2listIter$^*$
              \end{tabular}
& 168 & 12     & 174       & 12   & 4/0/1  & 22.93
                                  & 0.08       & 0.58        & 0.50 \\
  \hline

\footnotesize \begin{tabular}[c]{@{}l@{}}
                  \textbf{glib/glist\_{DLL}} (10): find, \textbf{\emph{free}}, index, last, \\
                  length, nth, nthData, position, prepend, reverse
                  \end{tabular}
      & 216 & 31     & 128       & 435(20)      & 9/1/0  & 403.13
  & 0       & 2.61     & 7.29    \\
  \hline

\footnotesize \begin{tabular}[c]{@{}l@{}}
                  \textbf{glib/glist\_{SLL}} (22): append, concat, copy, \textbf{\emph{delLink}}, find, \textbf{\emph{free}}, index, \\
                  insertAtPos, \emph{insertBefore}, \emph{insertSorted}, last, length, nth, nthData, \\
                  position, prepend, rm, rmAll, rmLink, reverse, sortMerge, \emph{sortReal}
                  \end{tabular}
  & 606 & 69     & 299     & 382(11)   & 17/5/0 & 879.35
  & 0.56       & 2.28     & 2.07    \\
  \hline

\footnotesize \begin{tabular}[c]{@{}l@{}}
                  \textbf{OpenBSD Queue} (6): init, insertAfter, insertHd, insertTl, \textbf{\emph{rmAfter}}, \textbf{\emph{rmHd}}
                  \end{tabular}
      & 105 & 12     & 12       & 27(4)      & 4/2/0  & 10.04
                          & 0.15       & 2.04     & 0.15    \\
  \hline

  \footnotesize \textbf{Memory Region} (1): memRegionDllOps
& 67  & 7      & 14       & 52        & 1/0/0  & 17.70
                          & 0.73          & 0.81     & 7.96    \\
  \hline

  \footnotesize \textbf{Binomial Heap} (2): \emph{findMin}, \emph{merge}
  & 117 & 8      & 54        & 89       & 0/2/0  & 76.56
                                & 1.39       & 0.90        & 9.15    \\
  \hline

  \footnotesize \begin{tabular}[c]{@{}l@{}}
                    \textbf{SV-COMP (Heap Programs)} (7): allocSlave, insertSlave,\\
                  createSlave, destroySlave, add, del, init
                  \end{tabular}
  & 119 & 16     & 34       & 71      & 7/0/0  & 58.17
  & 0.24       & 1.66     & 3.41    \\
  \hline

  \footnotesize \begin{tabular}[c]{@{}l@{}}
                  \textbf{GRASShopper\_SLL (Iterative)} (8): concat, copy, \textbf{\emph{dispose}}, filter, insert, \textbf{\emph{rm}}, \\
                  reverse, traverse
                  \end{tabular}
  & 193 & 27     & 111       & 98(9)     & 6/2/0 & 71.03
                          & 0.17       & 2.72     & 1.15    \\
  \hline

  \footnotesize \begin{tabular}[c]{@{}l@{}}
                  \textbf{GRASShopper\_SLL (Recursive)} (8): concat, copy, \textbf{\emph{dispose}}, filter, insert, \textbf{\emph{rm}}, \\
                  reverse, traverse
                  \end{tabular}
  & 173 & 24     & 118       & 40(3)       & 6/2/0 & 30.94
                          & 0.28       & 2.00     & 1.1    \\
  \hline

  \footnotesize \begin{tabular}[c]{@{}l@{}}
                  \textbf{GRASShopper\_DLL} (8): concat, copy, \textbf{\emph{dispose}}, filter$^{\dagger}$, insert, \textbf{\emph{rm}}, \\
                  reverse, traverse
                  \end{tabular}
  & 209 & 24     & 108       & 638(20)    & 5/2/1 & 803.58
                          & 0.04       & 2.95     & 8.50    \\
  \hline

\footnotesize \begin{tabular}[c]{@{}l@{}}
                  \textbf{GRASShopper\_SortedList} (14): concat, copy, \textbf{\emph{dispose}}, filter, insert, reverse,  \\
                \textbf{\emph{rm}}, split, traverse, merge,  doubleAll, pairwiseSum, insertionSort$^{\dagger}$, mergeSort$^*$
                  \end{tabular}
  & 394 & 43     & 195       & 222(1)     & 10/2/2 & 160.1
  & 1.04       & 2.27     & 4.29    \\
  \hline

\footnotesize \begin{tabular}[c]{@{}l@{}}
                \textbf{AFWP\_{SLL}} (11): create, \textbf{\emph{delAll}}, find, last, reverse, rotate, swap, insert, del$^{\dagger}$, \\
                \emph{filter}, \emph{merge}
                  \end{tabular}
  & 264 & 25     & 89       & 94(11)   & 7/3/1  & 71.04
  & 0.18      & 1.73     & 1.85    \\
  \hline

  \footnotesize  \textbf{AFWP\_{DLL}} (2): dll\_fix, dll\_splice
  & 40  & 5      & 16        & 133       & 2/0/0  & 75.51
                          & 0.02       & 2.96     & 6.67    \\
  \hline

  \footnotesize \textbf{Cyclist} (4): \emph{aplas-stack}, \emph{composite4}, \emph{iter}, schorr-waite
  & 506 & 32     & 360      & 132      & 1/3/0 & 165.26
                          & 0.27      & 0.63    & 0.67    \\
  \hline
\end{tabular}
\end{table*}

Table~\ref{tab:tab1} shows these 157 programs (the last row shows the 4 programs from~\cite{brotherston:popl16}).
Column \textbf{Programs} lists the programs, categorized by data structures that they use.
Column \textbf{LoC} shows the total lines of code of these programs.
For example, the first row lists 8 programs that use standard singly-link lists (SLL) and have in total 168 LoC.
In total, we have 157 programs in 22 categories with 4649 lines of C code.

\subsection{Setup}
\label{sec:setup}

For each program, we obtain traces, i.e., stack-heap models, to infer invariants at program entrances for preconditions,  at loop entrances for loop invariants, and at program exits for postconditions (we systematically obtain traces at each return statement in a program).
We use LLDB to set breakpoints at these locations to collect traces.

To obtain traces, we run each program on empty and randomly generated data structure inputs of a fixed size of 10.
For example, for the \code{concat} program in Figure~\ref{fig:concat} that takes as input 2 doubly-linked lists, we generate 3 inputs consisting of a nil list and two randomly generated doubly-linked lists $a,b$ of size 10.
Then we run \code{concat} over all input combinations, e.g., \code{(nil,a), (nil,b), (a,b)}, \dots.
Although these inputs are random and relatively small (size 10), the benchmark programs often modify and loop over data (e.g., as in \code{concat}), allowing us to generate sufficient and diverse traces.

For each category shown in Table~\ref{tab:tab1}, we adopt the predicate definitions given for that data from the benchmark programs, e.g., all programs \textbf{DLL} use the $\Dll$ inductive predicate shown in Section~\ref{sec:motiv}.
The shape and complexity of these predicates vary, e.g., $\Dll$ has 4 parameters, 1 singleton predicate, and 1 inductive predicate, and the \code{treeSeg} predicate has 2 parameters, 2 singletons, and 4 inductive predicates.

Programs in several categories such as \code{SV-COMP} and \code{Cyclist} use complex \emph{nested} data structures, which are data structures whose fields are other data structures.
The predicates for these data structures involve multiple predicates that are quite complex, e.g., the \code{iter} predicate has 10 parameters, 5 singleton and 6 inductive predicates.

\subsection{Results}
\label{sec:res}

Table~\ref{tab:tab1} shows our results.
Columns \textbf{iLocs}, \textbf{Traces}, and \textbf{Invs} lists, for programs in each category, the total number of target locations, obtained traces, and generated invariants, respectively.
Column \textbf{Invs} also lists the number of spurious invariants in parentheses (rows with no such parentheses have no spurious invariants).
Finally, column \textbf{Time} lists the total analysis time in seconds (including program execution, trace collection, and invariant inference).

For several programs, we were not able to obtain traces at considered locations using random inputs and thus could not infer invariants at those locations.
Column \textbf{A/S/X} shows the number of programs where we obtained traces at all considered locations (A), obtained traces for some locations or inferred spurious results (S), and could not obtain traces or invariants at some locations (X).
For example, for the 5 programs using binary search trees, we obtained traces at all considered locations in 2 programs, obtained traces at some locations in 2 programs, and could not obtain any traces in one program (\code{quicksort}).

The last three columns in the table give additional details about the generated invariants.
Columns \textbf{Single}, \textbf{Pred}, and \textbf{Pure} list the average
numbers of singleton predicates (e.g., $\formPtr{\code{node}}{x}{nx,pr}$),
inductive predicates (e.g., $\Dll$), and pure equalities (e.g., $x=\res$) found
in the invariants, respectively.

In total, {\tool} generated 3214 invariants in 487 target locations (average 6.60 invariants per location).
These invariants consists of 309 preconditions, 2442 postconditions and 463 loop invariants.
The total run time of {\tool} is 3866.06s for 149 programs\footnote{We exclude the 5 buggy programs that produce no traces and 3 programs that cause Z3 to time out.} (average 25.95s per program and 1.2s per invariant).
The time to run the program and collecting traces is negligible (about a second for all programs).

Out of 157 programs, we were not able to obtain any traces for 5 programs (marked with $^*$ in the table).
These programs contain bugs that immediately result in runtime errors such as segmentation faults (thus we obtained no traces and inferred no invariants).
For 15 programs (italic text), we could not reach certain \code{return} branches using random inputs and thus were not able to obtain traces or infer invariants at those locations. 
For 3 programs (marked with $\dagger$), we were able to generate pre and postconditions, but not loop invariants.
For these programs we hit loops more frequently than program entrance/exit points and thus obtained many traces for loops.
Checking generated formulae over many traces is expensive (Section~\ref{sec:complexity}) and appears to cause Z3 to stop responding.
Finally, for 17 programs (bold text), we obtained \emph{invalid} traces and therefore generated \emph{spurious} invariants.
This is an interesting behavior of running C programs and the LLDB debugger: a \code{free(x)} statement does not immediately free the pointer \code{x} so LLDB still observes (now invalid) heap values of \code{x} in the execution traces.
Thus we conservatively consider all generated invariants depending on these traces spurious and report them in Table~\ref{tab:tab1}.

For other programs (and those where we only obtain traces at certain locations), we manually analyzed and compared {\tool}'s generated invariants to documented ones.
First, we found that \emph{all} generated invariants are correct, i.e., they are true invariants at the considered locations.
Thus, the spurious results reported in Table~\ref{tab:tab1} are only those caused by invalid traces as described above.
Second, our results either matched (syntactically or semantically equivalent) or, in many cases, were
stronger than the documented invariants.
For example, for \code{SLL/reverse}, we inferred the documented postcondition $\formPred{\code{sll}}{\res}$ and the additional constraints $x\,{=}\,\nil \wedge x\,{=}\,\textsf{tmp}$ showing that the header of the input list $x$ becomes the tail of the resulting list.
In many similar cases, we achieved stronger results by inferring both the expected invariants and additional equalities.

A potential reason for these sound results is because {\tool} only infers shape properties using inductive predicates and pure equalities.
These properties have strict patterns and thus a property that holds for the observed traces will likely hold for others.
We also do not consider general disjunctive invariants or numerical relations (e.g., only check equivalences among memory addresses and do not consider other relationships such as the address of \code{x} is greater than that of \code{y}).
Existing numerical invariant studies~\cite{nguyen:tosem14:dig,nguyen:icse14:mpp} have shown that dynamic analysis often produces many spurious invariants involving disjunctions and general inequalities.

Although we tried our best to carefully check all generated results, the process of checking many complex SL invariants manually is time-consuming and difficult.
In future work, we will use an automatic verifier that supports SL formulae to check {\tool}'s invariants (see additional details in Section~\ref{sec:related}).
Moreover, we might be able to leverage test-input generation techniques, e.g., symbolic execution with
lazy-initialization~\cite{khurshid:tacas03:lazyinit} or SL predicates~\cite{web:jsf}, to construct smart inputs, which can explore hard-to-reach program paths to infer better invariants.

\subsection{Uses of Inferred Invariants}
Dynamically inferred invariants can help users understand programs (e.g., discovering loop invariants, pre and postconditions for unknown programs) and gain confidence about expected properties (e.g., the generated invariants met the expectation).
They can also be used to catch regression bugs: the user instruments these invariants as assertions in code to detect changes that break these assertions when the program run\footnote{The work in~\cite{nguyen:vmcai08:runtimecheckingsl} shows how to encode SL formulae, which contain non-standard operations such as $^*$, to executable functions that can be used assertions in code to enable run-time checking.}.
Existing works also list many other uses of dynamic invariants including documentation, complexity analysis, fault localization, and bug repair~\cite{ernst:scp07:daikon,le:issta16:faultlocinvs,perkins:sosp09:clearview,nguyen:ase17:syminfer}.
Below we show two concrete uses of {\tool}'s SL invariants.

\paragraph{Explaining Bugs}
Although {\tool} cannot generate inputs to reach a buggy location, it can, when given such inputs, discover useful invariants to alert and help the developer analyze that bug.
For \code{Red-black Tree/insert}, we obtained an invariant that appears too ``simple''.
Manual inspection showed that the inferred invariant is indeed correct: the program always crashes after the first iteration, thus the inferred invariant only captures a portion of data operated during the first iteration.
For \code{glib/glist\_SLL/sortMerge}, {\tool} reported an unexpected postcondition stating that the result is always null.
Manual inspection revealed this is correct and is due to a (typo) bug in the program that returns \code{list\_next} instead of \code{list->next}.
For \code{AFWP/dll\_fix.c},
the expected loop invariant is $\exists u_1, u_2, u_3, u_4.\,
\formPred{\code{sll}}{i} \sep \formPred{\Dll}{j, u_1, k, u_2} \sep
\formPred{\Dll}{k, u_3, u_4, \nil}$, but {\tool} returned
$\formPred{\code{sll}}{i} \wedge i\,{=}\,h \wedge k\,{=}\,j \wedge
k\,{=}\,\nil$. Thus, the expected invariant shows that $k$ can be non-{$\nil$}, but
{\tool}'s invariant shows the opposite.
Manual inspection showed a (potentially seeded) bug, where a guard checking for
$k\,{=}\,\nil$ was commented out. Indeed, with this guard uncommented, {\tool}
inferred the expected invariant.

\paragraph{Identifying Spurious Warnings}{\tool}'s invariants can help check results from static analyzers, e.g., to understand and gain confidence about reported results or detect potential problems.
The FBInfer tool mentioned in Section~\ref{sec:intro} is a well-known SL static analyzer that produces warnings for memory safety bugs for iOS and Android apps.
However, FBInfer can produce spurious (false positive) warnings.
For example, when analyzing the correct version of the mentioned \code{glib/glist\_SLL/sortMerge} program, FBInfer reported a memory leak after the assignment \code{l->next\;{=}\;NULL;} at the end of a loop because it thinks that \code{l->next} is not reachable.
However, {\tool}'s inferred invariants at that location showed \code{l->next} is a valid alias to other pointer variables and reachable.
Manual inspection confirmed that {\tool}'s generated invariants are correct and the program has no memory leak at that location.
We applied the same technique and found similar spurious warnings from FBInfer for 7 other programs\footnote{Programs with spurious warnings: \code{merge} in \code{Binomial\_Heap}, \code{delBack} in \code{Circular\_List}, \code{copy} in \code{Grasshopper\_SLL(Rec)}, \code{insert} in \code{GRASShopper\_DLL}, \code{GRASShoper\_SLL(Iter)}, \code{GRASShoper\_SortedList}, and \code{Grasshopper\_SLL(Rec)}.}.

Note that FBInfer also reported an error at another location of \code{sortMerge}.
However, this time, {\tool}'s invariants confirmed the warning and even revealed that the error is caused by a dangling pointer.

\subsection{Comparing to the S2 Static Analyzer}
\label{sec:compares2}
\begin{table}
\centering
\caption{Comparing SLING to S2.}
\label{tab:tab2}
\setlength\tabcolsep{2pt}
\footnotesize
\begin{tabular}{|l|c|c|c|c|c|}
\hline
\textbf{Programs} & \textbf{Total} & \textbf{Both} & \textbf{S2} & \textbf{SLING}  & \textbf{Neither} \\
 \hline

\footnotesize \textbf{SLL}                                  		 	& 9 & 8 & 0 & 1 & 0 \\
  \hline

\footnotesize \textbf{Sorted List}                         		& 14 & 6 & 0 & 6 & 2 \\
  \hline

\footnotesize \textbf{DLL}                                   			& 13 & 0 & 0 & 13 & 0 \\
  \hline

\footnotesize \textbf{Circular List}                       			& 4 & 0 & 0 & 2 & 2 \\
  \hline

\footnotesize \textbf{Binary Search Tree}           			& 6 & 1 & 1 & 2 & 2 \\
  \hline

\footnotesize \textbf{AVL Tree}                          			& 4 & 0 & 0 & 2 & 2 \\
  \hline

  \footnotesize \textbf{Priority Tree}                   			& 4 & 1 & 1 & 1 & 1 \\
  \hline

  \footnotesize \textbf{Red-black Tree}                		& 2 & 0 & 0 & 0 & 2 \\
  \hline

  \footnotesize \textbf{Tree Traversal}                			& 6 & 3 & 0 & 2 & 1 \\
  \hline

  \footnotesize \textbf{glib/glist\_DLL}                  			& 19 & 0 & 0 & 18 & 1 \\
  \hline

  \footnotesize \textbf{glib/glist\_SLL}                  			& 40 & 6 & 0 & 29 & 5 \\
  \hline

  \footnotesize \textbf{OpenBSD Queue}           			& 6 & 0 & 0 & 4 & 2 \\
  \hline

  \footnotesize \textbf{Memory Region}             			 & 3 & 1 & 0 & 2 & 0 \\
  \hline

  \footnotesize \textbf{Binomial Heap}                			& 2 & 0 & 1 & 0 & 1 \\
  \hline

    \footnotesize \textbf{SV-COMP } 		                      & 9 & 0 & 0 & 9 & 0 \\
  \hline

    \footnotesize \textbf{GRASShopper\_SLL (Iter)}  	& 16 & 2 & 0 & 12 & 2 \\
  \hline

    \footnotesize \textbf{GRASShopper\_SLL (Rec)}  	& 8 & 3 & 2 & 3 & 0 \\
  \hline

    \footnotesize \textbf{GRASShopper\_DLL }  			& 16 & 0 & 0 & 13 & 3 \\
  \hline

    \footnotesize \textbf{GRASShopper\_SortedList}  		& 29 & 1 & 0 & 24 & 4 \\
  \hline

    \footnotesize \textbf{AFWP\_SLL}  					& 20 & 1 & 0 & 15 & 4 \\
  \hline

    \footnotesize \textbf{AFWP\_DLL }  				& 3 & 0 & 0 & 3 & 0 \\
  \hline

    \footnotesize \textbf{Cyclist}  						& 4 & 0 & 0 & 1 & 3 \\
  \hline
  \hline
   \footnotesize \textbf{Total Sum}  						&  237 & 33 & 5 &  162 &  37\\
  \hline
\end{tabular}
\end{table}

We compare {\tool} to the static tool S2~\cite{le:cav14:s2}, which uses the state-of-the-art \emph{bi-abduction} technique~\cite{calcagno:jacm11} in SL to generate invariants proving memory safety of C programs, e.g., no null pointer dereferencing and memory leaks.
In addition to memory checking, S2 attempts to find strongest specifications consisting of pre and postconditions for heap programs.

We compare S2 to {\tool} using the same C benchmark programs\footnote{Several of these programs, e.g., those in \code{SLL}, \code{DLL}, and Binary trees, are also used in~\cite{le:cav14:s2} to evaluate S2's capability of proving memory safety.} listed in Table~\ref{tab:tab1}.
S2 only supports shape invariants, thus we only compare shape invariants generated by the two tools and ignore the pure invariants generated by {\tool}.
Moreover, S2 does not infer invariants at arbitrary locations like {\tool}, instead it attempts to find complete specifications (involving both pre and postconditions) and loop invariants.
Thus, we do not consider invariants generated at individual locations as shown in Table~\ref{tab:tab1} and instead consider specifications as a whole and loop invariants. Note that each program has a specification but only programs with loops have loop invariants.
As with \tool, we manual analyze the results of S2 and compare them to the documented invariants\footnote{We use these documented, verified invariants as ``ground truths'' and check if S2's results can match them. It is possible that these invariants are incomplete and not the ``best'' and thus non-matching results from S2 do not necessarily mean they are worse than these invariants.}.

Table~\ref{tab:tab2} shows the comparison results.
Column \textbf{Programs} lists the program categories, similarly to those listed in Table~\ref{tab:tab1}.
Column \textbf{Total} lists the number of documented properties consisting of specifications and loop invariants for the programs in the corresponding category.
The next four columns list the respective numbers of properties that \textbf{Both} tools can generate,
\textbf{S2} can generate but SLING cannot, \textbf{\tool} can generate but S2 cannot, and \textbf{Neither} tools can generate.
For example,
the 10 programs in \code{Sorted List} have 14 properties (10 specifications and 4 loop invariants), from which there are 6 properties that both tool found, none that only S2 found, 6 that only {\tool} found, and 2 that both fail to find.
Finally, S2 takes less than a second for all but the 4 \code{concat} programs in the \code{GRASShopper} categories, from which S2 appears to stop responding.

The last row of Table~\ref{tab:tab2} summarizes the results.
First, both tools discovered and failed about 14\% and 15\% of the properties, respectively.
Properties found by both tools are from simple recursive programs with singly-linked lists and trees.
For programs containing properties that neither tool found, we observed no patterns and different failure reasons, e.g., for \code{SLL/quicksort},  S2 did not produce any specification while {\tool} inferred no properties due the program crashed and produced no traces.
Next, S2 found 5 properties that {\tool} did not.
These properties are mostly in programs where {\tool} obtained incomplete or spurious results due to lack of traces, e.g., \code{binary\_search\_tree/find\_rec.c}, or the ``free'' problem, e.g., \code{GRASShopper/rec/dispose.c} (details in  Section~\ref{sec:res}).
Finally, {\tool} found many properties that S2 did not (162/237).
For these properties, which often come from complex programs with rich data structures, S2 either completely failed to produce them or produced much weaker than expected results.

In summary, {\tool} found more documented invariants comparing to S2.
We find this result encouraging as it shows the competitiveness of {\tool} to static analyzers.

\section{Related Work}
\label{sec:related}

{\tool} is inspired by the well-known dynamic invariant tool Daikon~\cite{ernst:scp07:daikon,ernst:tse01:dynamically}.
Daikon comes with a large list of invariant templates and predicates, tests them against program traces,  removes those that fail, and reports the remains as candidate invariants.
Recently, several techniques (such as PIE~\cite{padhi:pldi16:pie},
ICE~\cite{garg:popl16:ice}, DIG~\cite{nguyen:tosem14:dig}, and
SymInfer~\cite{nguyen:ase17:syminfer}) have been developed to infer numerical invariants using a hybrid approach that
dynamically infers candidate invariants and then statically checks them against the program code.
These approaches do not consider SL invariants for memory shape analysis.

Static program analysis in SL has rapidly gained adoption from both academia and industry in the past decade.
MemCAD~\cite{illous:nfm7:memcad} and THOR~\cite{magill:popl10:thor,magill:cav08:thor} reason about shape and numerical properties of programs, but generate invariants for a restricted language of list manipulating programs~\cite{magill:space06:inferring}.
FBInfer~\cite{calcagno:nasafm:moving,fb:web:fbinfer} uses bi-abduction to generate invariants to detect real memory bugs, but only supports simple structures (e.g., linked lists) and restricted language features (e.g., no arithmetic).
CABER~\cite{brotherston:sas14:caber} and S2 (described in Section~\ref{sec:compares2}) also use bi-abduction to offer more general, but more expensive, approaches to infer shape properties.
These tools do not consider invariants at arbitrary locations: CABER only analyzes preconditions and S2 only infers pre and postconditions.

The data-driven tool DOrder~\cite{zhu:pldi16:dorder} generates specifications for data structures in OCaml.  Given the definition of a data structure, DOrder generates predicates capturing shape and ordering relations among data (e.g., element $x$ is reachable from element $y$ or the value $x$ appears in the left subtree of a node containing the value $y$), learns specifications from predicates and in/output data, and verifies specifications using a refinement type system.
{\tool} makes an orthogonal contribution in finding a different form of shape properties to express sharing and aliasing information, e.g., nodes $x$ and $y$ point to (sub)trees in separate heaps.

The tool Locust~\cite{brockschmidt:sas17:locust} hybridizes dynamic and static analyses to infer SL invariants for programs written in a restricted language.
To infer an invariant, Locust expands the syntax of an SL formula using a machine learning model trained from a large set of data.
Locust iteratively refines inferred invariants using counterexamples obtained by the Grasshopper static verifier~\cite{piskac:tacas14:grasshopper}.
Locust is mainly evaluated on example programs with singly-linked lists and binary trees and does not support more complex data structures (e.g., Locust does not support doubly-linked list and returns no results when applied to our \code{concat} example).
The tool also relies on an expensive training process over large data sets. 

Finally, automatic verification tools such as HIP~\cite{chin:scp1212}, Grasshopper~\cite{piskac:tacas14:grasshopper}, Verifast~\cite{jacobs:nfm11:verifast}, and VCDryad~\cite{edgar:pldi14:naturalproofs} can prove given
SL specifications and invariants in heap-based programs.
In future work, we intend to use these tools to
automatically check {\tool}'s inferred invariants.

\section{Conclusion}
We introduce a new dynamic analysis to infer SL invariants for heap-manipulating programs.
The approach is based on the insight that the heap at a program location can be partitioned into disjoint regions reachable from various stack variables and that these regions can be modeled by atomic SL formulae.
Moreover, these formulae can be dynamically inferred and then combined using separating conjunction.

We present {\tool}, a tool that implements these ideas to generate SL invariants at arbitrary program locations.
{\tool} has several technical details including finding and using boundary variables instantiate predicates, using an SL model checker to compute both an instantiation for existential variables and the residual heap, and using the frame rule to validate inferred specifications.

Preliminary results on a large set of nontrivial programs show that {\tool} is effective in discovering useful invariants describing operations over a wide variety of data structures.
We believe that {\tool} takes an important step in broadening the space of properties about heap programs that can be dynamically inferred and exposes opportunities for researchers to exploit new dynamic SL invariant analyses.

\section*{Acknowledgments} We thank our shepherd Nadia Polikarpova and the anonymous reviewers for their feedback. The first author was supported by the Office of Naval Research award N000141712787.

\bibliography{paper}


\begin{thebibliography}{52}


\ifx \showCODEN    \undefined \def \showCODEN     #1{\unskip}     \fi
\ifx \showDOI      \undefined \def \showDOI       #1{#1}\fi
\ifx \showISBNx    \undefined \def \showISBNx     #1{\unskip}     \fi
\ifx \showISBNxiii \undefined \def \showISBNxiii  #1{\unskip}     \fi
\ifx \showISSN     \undefined \def \showISSN      #1{\unskip}     \fi
\ifx \showLCCN     \undefined \def \showLCCN      #1{\unskip}     \fi
\ifx \shownote     \undefined \def \shownote      #1{#1}          \fi
\ifx \showarticletitle \undefined \def \showarticletitle #1{#1}   \fi
\ifx \showURL      \undefined \def \showURL       {\relax}        \fi
\providecommand\bibfield[2]{#2}
\providecommand\bibinfo[2]{#2}
\providecommand\natexlab[1]{#1}
\providecommand\showeprint[2][]{arXiv:#2}

\bibitem[\protect\citeauthoryear{B~Le, Lo, Le~Goues, and Grunske}{B~Le
  et~al\mbox{.}}{2016}]%
        {le:issta16:faultlocinvs}
\bibfield{author}{\bibinfo{person}{Tien-Duy B~Le}, \bibinfo{person}{David Lo},
  \bibinfo{person}{Claire Le~Goues}, {and} \bibinfo{person}{Lars Grunske}.}
  \bibinfo{year}{2016}\natexlab{}.
\newblock \showarticletitle{A learning-to-rank based fault localization
  approach using likely invariants}. In \bibinfo{booktitle}{\emph{ISSTA}}. ACM,
  \bibinfo{pages}{177--188}.
\newblock


\bibitem[\protect\citeauthoryear{Ball and Rajamani}{Ball and Rajamani}{2002}]%
        {ball:ifm04:slam}
\bibfield{author}{\bibinfo{person}{Thomas Ball} {and}
  \bibinfo{person}{Sriram~K. Rajamani}.} \bibinfo{year}{2002}\natexlab{}.
\newblock \showarticletitle{{The SLAM Project: Debugging System Software via
  Static Analysis}}. In \bibinfo{booktitle}{\emph{POPL}}.
  \bibinfo{pages}{1--3}.
\newblock
\showISBNx{1-58113-450-9}


\bibitem[\protect\citeauthoryear{Berdine, Calcagno, and O'Hearn}{Berdine
  et~al\mbox{.}}{2004}]%
        {berdine:fsttcs04}
\bibfield{author}{\bibinfo{person}{Josh Berdine}, \bibinfo{person}{Cristiano
  Calcagno}, {and} \bibinfo{person}{Peter~W. O'Hearn}.}
  \bibinfo{year}{2004}\natexlab{}.
\newblock \showarticletitle{{A Decidable Fragment of Separation Logic}}. In
  \bibinfo{booktitle}{\emph{Foundations of Software Technology and Theoretical
  Computer Science}}. \bibinfo{pages}{97--109}.
\newblock


\bibitem[\protect\citeauthoryear{Beyer}{Beyer}{2017}]%
        {beyer:tacas17:svcomp}
\bibfield{author}{\bibinfo{person}{Dirk Beyer}.}
  \bibinfo{year}{2017}\natexlab{}.
\newblock \showarticletitle{Software Verification with Validation of Results}.
  In \bibinfo{booktitle}{\emph{TACAS}}. \bibinfo{pages}{331--349}.
\newblock


\bibitem[\protect\citeauthoryear{Brockschmidt, Chen, Kohli, Krishna, and
  Tarlow}{Brockschmidt et~al\mbox{.}}{2017}]%
        {brockschmidt:sas17:locust}
\bibfield{author}{\bibinfo{person}{Marc Brockschmidt}, \bibinfo{person}{Yuxin
  Chen}, \bibinfo{person}{Pushmeet Kohli}, \bibinfo{person}{Siddharth Krishna},
  {and} \bibinfo{person}{Daniel Tarlow}.} \bibinfo{year}{2017}\natexlab{}.
\newblock \showarticletitle{Learning Shape Analysis}. In
  \bibinfo{booktitle}{\emph{Static Analysis Symposium}}.
  \bibinfo{pages}{66--87}.
\newblock


\bibitem[\protect\citeauthoryear{Brotherston and Gorogiannis}{Brotherston and
  Gorogiannis}{2014}]%
        {brotherston:sas14:caber}
\bibfield{author}{\bibinfo{person}{James Brotherston} {and}
  \bibinfo{person}{Nikos Gorogiannis}.} \bibinfo{year}{2014}\natexlab{}.
\newblock \showarticletitle{{Cyclic Abduction of Inductively Defined Safety and
  Termination Preconditions}}. In \bibinfo{booktitle}{\emph{Static Analysis}}.
  \bibinfo{pages}{68--84}.
\newblock


\bibitem[\protect\citeauthoryear{Brotherston, Gorogiannis, Kanovich, and
  Rowe}{Brotherston et~al\mbox{.}}{2016}]%
        {brotherston:popl16}
\bibfield{author}{\bibinfo{person}{James Brotherston}, \bibinfo{person}{Nikos
  Gorogiannis}, \bibinfo{person}{Max~I. Kanovich}, {and}
  \bibinfo{person}{Reuben Rowe}.} \bibinfo{year}{2016}\natexlab{}.
\newblock \showarticletitle{{Model checking for symbolic-heap separation logic
  with inductive predicates}}. In \bibinfo{booktitle}{\emph{POPL}}.
  \bibinfo{pages}{84--96}.
\newblock


\bibitem[\protect\citeauthoryear{Calcagno, Distefano, Dubreil, Gabi,
  Hooimeijer, Luca, O'Hearn, Papakonstantinou, Purbrick, and
  Rodriguez}{Calcagno et~al\mbox{.}}{2015}]%
        {calcagno:nasafm:moving}
\bibfield{author}{\bibinfo{person}{Cristiano Calcagno}, \bibinfo{person}{Dino
  Distefano}, \bibinfo{person}{J{\'{e}}r{\'{e}}my Dubreil},
  \bibinfo{person}{Dominik Gabi}, \bibinfo{person}{Pieter Hooimeijer},
  \bibinfo{person}{Martino Luca}, \bibinfo{person}{Peter~W. O'Hearn},
  \bibinfo{person}{Irene Papakonstantinou}, \bibinfo{person}{Jim Purbrick},
  {and} \bibinfo{person}{Dulma Rodriguez}.} \bibinfo{year}{2015}\natexlab{}.
\newblock \showarticletitle{{Moving Fast with Software Verification}}. In
  \bibinfo{booktitle}{\emph{{NASA Formal Methods}}}. \bibinfo{pages}{3--11}.
\newblock


\bibitem[\protect\citeauthoryear{Calcagno, Distefano, O'Hearn, and
  Yang}{Calcagno et~al\mbox{.}}{2011}]%
        {calcagno:jacm11}
\bibfield{author}{\bibinfo{person}{Cristiano Calcagno}, \bibinfo{person}{Dino
  Distefano}, \bibinfo{person}{Peter~W. O'Hearn}, {and}
  \bibinfo{person}{Hongseok Yang}.} \bibinfo{year}{2011}\natexlab{}.
\newblock \showarticletitle{{Compositional Shape Analysis by Means of
  Bi-Abduction}}.
\newblock \bibinfo{journal}{\emph{J. {ACM}}} \bibinfo{volume}{58},
  \bibinfo{number}{6} (\bibinfo{year}{2011}), \bibinfo{pages}{26:1--26:66}.
\newblock


\bibitem[\protect\citeauthoryear{Chin, David, Nguyen, and Qin}{Chin
  et~al\mbox{.}}{2012}]%
        {chin:scp1212}
\bibfield{author}{\bibinfo{person}{Wei{-}Ngan Chin}, \bibinfo{person}{Cristina
  David}, \bibinfo{person}{Huu~Hai Nguyen}, {and} \bibinfo{person}{Shengchao
  Qin}.} \bibinfo{year}{2012}\natexlab{}.
\newblock \showarticletitle{Automated verification of shape, size and bag
  properties via user-defined predicates in separation logic}.
\newblock \bibinfo{journal}{\emph{Sci. Comput. Program.}} \bibinfo{volume}{77},
  \bibinfo{number}{9} (\bibinfo{year}{2012}), \bibinfo{pages}{1006--1036}.
\newblock


\bibitem[\protect\citeauthoryear{Cook, Haase, Ouaknine, Parkinson, and
  Worrell}{Cook et~al\mbox{.}}{2011}]%
        {cook:concur11}
\bibfield{author}{\bibinfo{person}{Byron Cook}, \bibinfo{person}{Christoph
  Haase}, \bibinfo{person}{Jo{\"{e}}l Ouaknine}, \bibinfo{person}{Matthew~J.
  Parkinson}, {and} \bibinfo{person}{James Worrell}.}
  \bibinfo{year}{2011}\natexlab{}.
\newblock \showarticletitle{{Tractable Reasoning in a Fragment of Separation
  Logic}}. In \bibinfo{booktitle}{\emph{CONCUR}}. \bibinfo{pages}{235--249}.
\newblock


\bibitem[\protect\citeauthoryear{De~Moura and Bj{\o}rner}{De~Moura and
  Bj{\o}rner}{2008}]%
        {demoura:tacas08:z3}
\bibfield{author}{\bibinfo{person}{Leonardo De~Moura} {and}
  \bibinfo{person}{Nikolaj Bj{\o}rner}.} \bibinfo{year}{2008}\natexlab{}.
\newblock \showarticletitle{Z3: {A}n efficient {SMT} solver}.
\newblock In \bibinfo{booktitle}{\emph{TACAS}}. \bibinfo{pages}{337--340}.
\newblock


\bibitem[\protect\citeauthoryear{Dijkstra}{Dijkstra}{1975}]%
        {dijkstra:cacm75:wp}
\bibfield{author}{\bibinfo{person}{Edsger~W. Dijkstra}.}
  \bibinfo{year}{1975}\natexlab{}.
\newblock \showarticletitle{Guarded commands, nondeterminacy and formal
  derivation of programs}.
\newblock \bibinfo{journal}{\emph{Commun. ACM}}  \bibinfo{volume}{18}
  (\bibinfo{year}{1975}), \bibinfo{pages}{453--457}.
\newblock
Issue 8.


\bibitem[\protect\citeauthoryear{Ernst, Cockrell, Griswold, and Notkin}{Ernst
  et~al\mbox{.}}{2001}]%
        {ernst:tse01:dynamically}
\bibfield{author}{\bibinfo{person}{Michael~D Ernst}, \bibinfo{person}{Jake
  Cockrell}, \bibinfo{person}{William~G Griswold}, {and} \bibinfo{person}{David
  Notkin}.} \bibinfo{year}{2001}\natexlab{}.
\newblock \showarticletitle{Dynamically discovering likely program invariants
  to support program evolution}.
\newblock \bibinfo{journal}{\emph{Transactions on Software Engineering}}
  \bibinfo{volume}{27}, \bibinfo{number}{2} (\bibinfo{year}{2001}),
  \bibinfo{pages}{99--123}.
\newblock


\bibitem[\protect\citeauthoryear{Ernst, Perkins, Guo, McCamant, Pacheco,
  Tschantz, and Xiao}{Ernst et~al\mbox{.}}{2007}]%
        {ernst:scp07:daikon}
\bibfield{author}{\bibinfo{person}{Michael~D. Ernst}, \bibinfo{person}{Jeff~H.
  Perkins}, \bibinfo{person}{Philip~J. Guo}, \bibinfo{person}{Stephen
  McCamant}, \bibinfo{person}{Carlos Pacheco}, \bibinfo{person}{Matthew~S.
  Tschantz}, {and} \bibinfo{person}{Chen Xiao}.}
  \bibinfo{year}{2007}\natexlab{}.
\newblock \showarticletitle{The {Daikon} system for dynamic detection of likely
  invariants}.
\newblock \bibinfo{journal}{\emph{Science of Computer Programming}}
  (\bibinfo{year}{2007}), \bibinfo{pages}{35--45}.
\newblock


\bibitem[\protect\citeauthoryear{Fava, Signoles, Lemerre, Sch{\"a}f, and
  Tiwari}{Fava et~al\mbox{.}}{2015}]%
        {fava:lpar15:gamifying}
\bibfield{author}{\bibinfo{person}{Daniel Fava}, \bibinfo{person}{Julien
  Signoles}, \bibinfo{person}{Matthieu Lemerre}, \bibinfo{person}{Martin
  Sch{\"a}f}, {and} \bibinfo{person}{Ashish Tiwari}.}
  \bibinfo{year}{2015}\natexlab{}.
\newblock \showarticletitle{Gamifying program analysis}. In
  \bibinfo{booktitle}{\emph{LPAR}}. Springer, \bibinfo{pages}{591--605}.
\newblock


\bibitem[\protect\citeauthoryear{FBInfer}{FBInfer}{2018}]%
        {fb:web:fbinfer}
FBInfer \bibinfo{year}{2018}\natexlab{}.
\newblock \bibinfo{title}{{The Infer Static Analyzer}}.
\newblock
\newblock
\newblock
\shownote{\url{http://fbinfer.com/}.}


\bibitem[\protect\citeauthoryear{Flanagan and Leino}{Flanagan and
  Leino}{2001}]%
        {flanagan:fm01:houdini}
\bibfield{author}{\bibinfo{person}{Cormac Flanagan} {and}
  \bibinfo{person}{K.~Rustan~M. Leino}.} \bibinfo{year}{2001}\natexlab{}.
\newblock \showarticletitle{Houdini, an Annotation Assistant for {ESC/Java}}.
  In \bibinfo{booktitle}{\emph{Formal Methods for Increasing Software
  Productivity}}. \bibinfo{pages}{500--517}.
\newblock


\bibitem[\protect\citeauthoryear{Garg, Neider, Madhusudan, and Roth}{Garg
  et~al\mbox{.}}{2016}]%
        {garg:popl16:ice}
\bibfield{author}{\bibinfo{person}{Pranav Garg}, \bibinfo{person}{Daniel
  Neider}, \bibinfo{person}{P. Madhusudan}, {and} \bibinfo{person}{Dan Roth}.}
  \bibinfo{year}{2016}\natexlab{}.
\newblock \showarticletitle{{Learning Invariants Using Decision Trees and
  Implication Counterexamples}}. In \bibinfo{booktitle}{\emph{POPL}}.
  \bibinfo{pages}{499--512}.
\newblock


\bibitem[\protect\citeauthoryear{GDB}{GDB}{2018}]%
        {web:gdb}
GDB \bibinfo{year}{2018}\natexlab{}.
\newblock \bibinfo{title}{{GDB: The GNU Project Debugger}}.
\newblock
\newblock
\newblock
\shownote{\url{https://www.gnu.org/software/gdb/}.}


\bibitem[\protect\citeauthoryear{Hoare}{Hoare}{1969}]%
        {hoare:69:axiomatic}
\bibfield{author}{\bibinfo{person}{Charles Antony~Richard Hoare}.}
  \bibinfo{year}{1969}\natexlab{}.
\newblock \showarticletitle{An axiomatic basis for computer programming}.
\newblock \bibinfo{journal}{\emph{Commun. ACM}} \bibinfo{volume}{12},
  \bibinfo{number}{10} (\bibinfo{year}{1969}), \bibinfo{pages}{576--580}.
\newblock


\bibitem[\protect\citeauthoryear{Illous, Lemerre, and Rival}{Illous
  et~al\mbox{.}}{2017}]%
        {illous:nfm7:memcad}
\bibfield{author}{\bibinfo{person}{Hugo Illous}, \bibinfo{person}{Matthieu
  Lemerre}, {and} \bibinfo{person}{Xavier Rival}.}
  \bibinfo{year}{2017}\natexlab{}.
\newblock \showarticletitle{A Relational Shape Abstract Domain}. In
  \bibinfo{booktitle}{\emph{{NASA} Formal Methods}}. \bibinfo{pages}{212--229}.
\newblock


\bibitem[\protect\citeauthoryear{Itzhaky, Banerjee, Immerman, Nanevski, and
  Sagiv}{Itzhaky et~al\mbox{.}}{2013}]%
        {itzhaky:cav13}
\bibfield{author}{\bibinfo{person}{Shachar Itzhaky}, \bibinfo{person}{Anindya
  Banerjee}, \bibinfo{person}{Neil Immerman}, \bibinfo{person}{Aleksandar
  Nanevski}, {and} \bibinfo{person}{Mooly Sagiv}.}
  \bibinfo{year}{2013}\natexlab{}.
\newblock \showarticletitle{Effectively-Propositional Reasoning about
  Reachability in Linked Data Structures}. In \bibinfo{booktitle}{\emph{CAV}}.
  \bibinfo{pages}{756--772}.
\newblock


\bibitem[\protect\citeauthoryear{Jacobs, Smans, Philippaerts, Vogels,
  Penninckx, and Piessens}{Jacobs et~al\mbox{.}}{2011}]%
        {jacobs:nfm11:verifast}
\bibfield{author}{\bibinfo{person}{Bart Jacobs}, \bibinfo{person}{Jan Smans},
  \bibinfo{person}{Pieter Philippaerts}, \bibinfo{person}{Fr{\'{e}}d{\'{e}}ric
  Vogels}, \bibinfo{person}{Willem Penninckx}, {and} \bibinfo{person}{Frank
  Piessens}.} \bibinfo{year}{2011}\natexlab{}.
\newblock \showarticletitle{{VeriFast: {A} Powerful, Sound, Predictable, Fast
  Verifier for {C} and Java}}. In \bibinfo{booktitle}{\emph{{NASA} Formal
  Methods}}. \bibinfo{pages}{41--55}.
\newblock


\bibitem[\protect\citeauthoryear{JDB}{JDB}{2018}]%
        {web:jdb}
JDB \bibinfo{year}{2018}\natexlab{}.
\newblock \bibinfo{title}{{jdb - The Java Debugger}}.
\newblock
\newblock
\newblock
\shownote{\url{https://docs.oracle.com/javase/8/docs/technotes/tools/windows/jdb.html}.}


\bibitem[\protect\citeauthoryear{Jones and Muchnick}{Jones and
  Muchnick}{1982}]%
        {jones:popl82}
\bibfield{author}{\bibinfo{person}{Neil~D. Jones} {and}
  \bibinfo{person}{Steven~S. Muchnick}.} \bibinfo{year}{1982}\natexlab{}.
\newblock \showarticletitle{{A Flexible Approach to Interprocedural Data Flow
  Analysis and Programs with Recursive Data Structures}}. In
  \bibinfo{booktitle}{\emph{POPL}}. \bibinfo{pages}{66--74}.
\newblock


\bibitem[\protect\citeauthoryear{JSF}{JSF}{2018}]%
        {web:jsf}
JSF \bibinfo{year}{2018}\natexlab{}.
\newblock \bibinfo{title}{{JSF: The Java StarFinder Symbolic Execution Tool}}.
\newblock
\newblock
\newblock
\shownote{\url{https://github.com/star-finder/jpf-star}.}


\bibitem[\protect\citeauthoryear{Khurshid, P{\u{a}}s{\u{a}}reanu, and
  Visser}{Khurshid et~al\mbox{.}}{2003}]%
        {khurshid:tacas03:lazyinit}
\bibfield{author}{\bibinfo{person}{Sarfraz Khurshid}, \bibinfo{person}{Corina~S
  P{\u{a}}s{\u{a}}reanu}, {and} \bibinfo{person}{Willem Visser}.}
  \bibinfo{year}{2003}\natexlab{}.
\newblock \showarticletitle{Generalized symbolic execution for model checking
  and testing}. In \bibinfo{booktitle}{\emph{TACAS}}. Springer,
  \bibinfo{pages}{553--568}.
\newblock


\bibitem[\protect\citeauthoryear{Le, Gherghina, Qin, and Chin}{Le
  et~al\mbox{.}}{2014}]%
        {le:cav14:s2}
\bibfield{author}{\bibinfo{person}{Quang~Loc Le}, \bibinfo{person}{Cristian
  Gherghina}, \bibinfo{person}{Shengchao Qin}, {and}
  \bibinfo{person}{Wei{-}Ngan Chin}.} \bibinfo{year}{2014}\natexlab{}.
\newblock \showarticletitle{Shape Analysis via Second-Order Bi-Abduction}. In
  \bibinfo{booktitle}{\emph{CAV}}. \bibinfo{pages}{52--68}.
\newblock


\bibitem[\protect\citeauthoryear{Leroy}{Leroy}{2006}]%
        {leroy:popl06:}
\bibfield{author}{\bibinfo{person}{Xavier Leroy}.}
  \bibinfo{year}{2006}\natexlab{}.
\newblock \showarticletitle{Formal certification of a compiler back-end or:
  programming a compiler with a proof assistant}. In
  \bibinfo{booktitle}{\emph{POPL}}. \bibinfo{pages}{42--54}.
\newblock


\bibitem[\protect\citeauthoryear{LLDB}{LLDB}{2018}]%
        {web:lldb}
LLDB \bibinfo{year}{2018}\natexlab{}.
\newblock \bibinfo{title}{The {LLDB} {D}ebugger}.
\newblock
\newblock
\newblock
\shownote{\url{https://lldb.llvm.org/}.}


\bibitem[\protect\citeauthoryear{Magill, Nanevski, Clarke, and Lee}{Magill
  et~al\mbox{.}}{2006}]%
        {magill:space06:inferring}
\bibfield{author}{\bibinfo{person}{Stephen Magill}, \bibinfo{person}{Aleksandar
  Nanevski}, \bibinfo{person}{Edmund Clarke}, {and} \bibinfo{person}{Peter
  Lee}.} \bibinfo{year}{2006}\natexlab{}.
\newblock \showarticletitle{Inferring invariants in separation logic for
  imperative list-processing programs}.
\newblock \bibinfo{journal}{\emph{SPACE}} \bibinfo{volume}{1},
  \bibinfo{number}{1} (\bibinfo{year}{2006}), \bibinfo{pages}{5--7}.
\newblock


\bibitem[\protect\citeauthoryear{Magill, Tsai, Lee, and Tsay}{Magill
  et~al\mbox{.}}{2008}]%
        {magill:cav08:thor}
\bibfield{author}{\bibinfo{person}{Stephen Magill},
  \bibinfo{person}{Ming{-}Hsien Tsai}, \bibinfo{person}{Peter Lee}, {and}
  \bibinfo{person}{Yih{-}Kuen Tsay}.} \bibinfo{year}{2008}\natexlab{}.
\newblock \showarticletitle{{THOR:} {A} Tool for Reasoning about Shape and
  Arithmetic}. In \bibinfo{booktitle}{\emph{CAV}}. \bibinfo{pages}{428--432}.
\newblock


\bibitem[\protect\citeauthoryear{Magill, Tsai, Lee, and Tsay}{Magill
  et~al\mbox{.}}{2010}]%
        {magill:popl10:thor}
\bibfield{author}{\bibinfo{person}{Stephen Magill},
  \bibinfo{person}{Ming{-}Hsien Tsai}, \bibinfo{person}{Peter Lee}, {and}
  \bibinfo{person}{Yih{-}Kuen Tsay}.} \bibinfo{year}{2010}\natexlab{}.
\newblock \showarticletitle{Automatic numeric abstractions for
  heap-manipulating programs}. In \bibinfo{booktitle}{\emph{POPL}}.
  \bibinfo{pages}{211--222}.
\newblock


\bibitem[\protect\citeauthoryear{Navarro~P{\'{e}}rez and
  Rybalchenko}{Navarro~P{\'{e}}rez and Rybalchenko}{2011}]%
        {perez:pldi11}
\bibfield{author}{\bibinfo{person}{Juan~Antonio Navarro~P{\'{e}}rez} {and}
  \bibinfo{person}{Andrey Rybalchenko}.} \bibinfo{year}{2011}\natexlab{}.
\newblock \showarticletitle{{Separation Logic + Superposition Calculus = Heap
  Theorem Prover}}. In \bibinfo{booktitle}{\emph{PLDI}}.
  \bibinfo{pages}{556--566}.
\newblock


\bibitem[\protect\citeauthoryear{Nguyen, Kuncak, and Chin}{Nguyen
  et~al\mbox{.}}{2008}]%
        {nguyen:vmcai08:runtimecheckingsl}
\bibfield{author}{\bibinfo{person}{Huu~Hai Nguyen}, \bibinfo{person}{Viktor
  Kuncak}, {and} \bibinfo{person}{Wei-Ngan Chin}.}
  \bibinfo{year}{2008}\natexlab{}.
\newblock \showarticletitle{Runtime Checking for Separation Logic}. In
  \bibinfo{booktitle}{\emph{VMCAI}}. \bibinfo{pages}{203--217}.
\newblock


\bibitem[\protect\citeauthoryear{Nguyen, Dwyer, and Visser}{Nguyen
  et~al\mbox{.}}{2017}]%
        {nguyen:ase17:syminfer}
\bibfield{author}{\bibinfo{person}{ThanhVu Nguyen}, \bibinfo{person}{Matthew
  Dwyer}, {and} \bibinfo{person}{William Visser}.}
  \bibinfo{year}{2017}\natexlab{}.
\newblock \showarticletitle{{SymInfer: Inferring Program Invariants using
  Symbolic States}}. In \bibinfo{booktitle}{\emph{ASE}}.
  \bibinfo{pages}{804--814}.
\newblock


\bibitem[\protect\citeauthoryear{Nguyen, Kapur, Weimer, and Forrest}{Nguyen
  et~al\mbox{.}}{2014a}]%
        {nguyen:tosem14:dig}
\bibfield{author}{\bibinfo{person}{ThanhVu Nguyen}, \bibinfo{person}{Deepak
  Kapur}, \bibinfo{person}{Westley Weimer}, {and} \bibinfo{person}{Stephanie
  Forrest}.} \bibinfo{year}{2014}\natexlab{a}.
\newblock \showarticletitle{{DIG: A Dynamic Invariant Generator for Polynomial
  and Array Invariants}}.
\newblock \bibinfo{journal}{\emph{Transactions on Software Engineering
  Methodology}} \bibinfo{volume}{23}, \bibinfo{number}{4}
  (\bibinfo{year}{2014}), \bibinfo{pages}{30:1--30:30}.
\newblock


\bibitem[\protect\citeauthoryear{Nguyen, Kapur, Weimer, and Forrest}{Nguyen
  et~al\mbox{.}}{2014b}]%
        {nguyen:icse14:mpp}
\bibfield{author}{\bibinfo{person}{ThanhVu Nguyen}, \bibinfo{person}{Deepak
  Kapur}, \bibinfo{person}{Westley Weimer}, {and} \bibinfo{person}{Stephanie
  Forrest}.} \bibinfo{year}{2014}\natexlab{b}.
\newblock \showarticletitle{{Using Dynamic Analysis to Generate Disjunctive
  Invariants.}}. In \bibinfo{booktitle}{\emph{ICSE}}.
  \bibinfo{pages}{608--619}.
\newblock


\bibitem[\protect\citeauthoryear{O'Hearn}{O'Hearn}{2016}]%
        {ohearn:web16:}
\bibfield{author}{\bibinfo{person}{Peter~W. O'Hearn}.}
  \bibinfo{year}{2016}\natexlab{}.
\newblock \bibinfo{title}{CurryOn '16 Talk: Move fast to fix more things}.
\newblock
\newblock


\bibitem[\protect\citeauthoryear{O'Hearn, Reynolds, and Yang}{O'Hearn
  et~al\mbox{.}}{2001}]%
        {ohearn:csl01:sl}
\bibfield{author}{\bibinfo{person}{Peter~W. O'Hearn}, \bibinfo{person}{John~C.
  Reynolds}, {and} \bibinfo{person}{Hongseok Yang}.}
  \bibinfo{year}{2001}\natexlab{}.
\newblock \showarticletitle{{Local Reasoning about Programs that Alter Data
  Structures}}. In \bibinfo{booktitle}{\emph{CSL}}. \bibinfo{pages}{1--19}.
\newblock


\bibitem[\protect\citeauthoryear{Padhi, Sharma, and Millstein}{Padhi
  et~al\mbox{.}}{2016}]%
        {padhi:pldi16:pie}
\bibfield{author}{\bibinfo{person}{Saswat Padhi}, \bibinfo{person}{Rahul
  Sharma}, {and} \bibinfo{person}{Todd Millstein}.}
  \bibinfo{year}{2016}\natexlab{}.
\newblock \showarticletitle{Data-driven Precondition Inference with Learned
  Features}. In \bibinfo{booktitle}{\emph{PLDI}}. \bibinfo{pages}{42--56}.
\newblock


\bibitem[\protect\citeauthoryear{PDB}{PDB}{2018}]%
        {web:pdb}
PDB \bibinfo{year}{2018}\natexlab{}.
\newblock \bibinfo{title}{{pdb - The Python Debugger}}.
\newblock
\newblock
\newblock
\shownote{\url{https://docs.python.org/2/library/pdb.html}.}


\bibitem[\protect\citeauthoryear{Pek, Qiu, and Madhusudan}{Pek
  et~al\mbox{.}}{2014}]%
        {edgar:pldi14:naturalproofs}
\bibfield{author}{\bibinfo{person}{Edgar Pek}, \bibinfo{person}{Xiaokang Qiu},
  {and} \bibinfo{person}{P. Madhusudan}.} \bibinfo{year}{2014}\natexlab{}.
\newblock \showarticletitle{Natural Proofs for Data Structure Manipulation in C
  Using Separation Logic}. In \bibinfo{booktitle}{\emph{PLDI}}.
  \bibinfo{pages}{440--451}.
\newblock
\showISBNx{978-1-4503-2784-8}


\bibitem[\protect\citeauthoryear{Perkins, Kim, Larsen, Amarasinghe, Bachrach,
  Carbin, Pacheco, Sherwood, Sidiroglou, Sullivan, Wong, Zibin, Ernst, and
  Rinard}{Perkins et~al\mbox{.}}{2009}]%
        {perkins:sosp09:clearview}
\bibfield{author}{\bibinfo{person}{Jeff~H. Perkins}, \bibinfo{person}{Sunghun
  Kim}, \bibinfo{person}{Sam Larsen}, \bibinfo{person}{Saman Amarasinghe},
  \bibinfo{person}{Jonathan Bachrach}, \bibinfo{person}{Michael Carbin},
  \bibinfo{person}{Carlos Pacheco}, \bibinfo{person}{Frank Sherwood},
  \bibinfo{person}{Stelios Sidiroglou}, \bibinfo{person}{Greg Sullivan},
  \bibinfo{person}{Weng-Fai Wong}, \bibinfo{person}{Yoav Zibin},
  \bibinfo{person}{Michael~D. Ernst}, {and} \bibinfo{person}{Martin Rinard}.}
  \bibinfo{year}{2009}\natexlab{}.
\newblock \showarticletitle{Automatically patching errors in deployed
  software}. In \bibinfo{booktitle}{\emph{Symposium on Operating Systems
  Principles}}. \bibinfo{pages}{87--102}.
\newblock


\bibitem[\protect\citeauthoryear{Piskac, Wies, and Zufferey}{Piskac
  et~al\mbox{.}}{2014}]%
        {piskac:tacas14:grasshopper}
\bibfield{author}{\bibinfo{person}{Ruzica Piskac}, \bibinfo{person}{Thomas
  Wies}, {and} \bibinfo{person}{Damien Zufferey}.}
  \bibinfo{year}{2014}\natexlab{}.
\newblock \showarticletitle{GRASShopper - Complete Heap Verification with Mixed
  Specifications}. In \bibinfo{booktitle}{\emph{TACAS}}.
  \bibinfo{pages}{124--139}.
\newblock


\bibitem[\protect\citeauthoryear{Reynolds}{Reynolds}{2002}]%
        {reynolds:lics02:sl}
\bibfield{author}{\bibinfo{person}{John~C. Reynolds}.}
  \bibinfo{year}{2002}\natexlab{}.
\newblock \showarticletitle{{Separation Logic: {A} Logic for Shared Mutable
  Data Structures}}. In \bibinfo{booktitle}{\emph{Symposium on Logic in
  Computer Science}}. \bibinfo{pages}{55--74}.
\newblock


\bibitem[\protect\citeauthoryear{Sagiv, Reps, and Wilhelm}{Sagiv
  et~al\mbox{.}}{2002}]%
        {sagiv:toplas2002:parametric}
\bibfield{author}{\bibinfo{person}{Mooly Sagiv}, \bibinfo{person}{Thomas Reps},
  {and} \bibinfo{person}{Reinhard Wilhelm}.} \bibinfo{year}{2002}\natexlab{}.
\newblock \showarticletitle{Parametric shape analysis via 3-valued logic}.
\newblock \bibinfo{journal}{\emph{TOPLAS}} \bibinfo{volume}{24},
  \bibinfo{number}{3} (\bibinfo{year}{2002}), \bibinfo{pages}{217--298}.
\newblock


\bibitem[\protect\citeauthoryear{Ta, Le, Khoo, and Chin}{Ta
  et~al\mbox{.}}{2016}]%
        {ta:fm16}
\bibfield{author}{\bibinfo{person}{Quang{-}Trung Ta},
  \bibinfo{person}{Ton~Chanh Le}, \bibinfo{person}{Siau{-}Cheng Khoo}, {and}
  \bibinfo{person}{Wei{-}Ngan Chin}.} \bibinfo{year}{2016}\natexlab{}.
\newblock \showarticletitle{Automated Mutual Explicit Induction Proof in
  Separation Logic}. In \bibinfo{booktitle}{\emph{FM}}.
  \bibinfo{pages}{659--676}.
\newblock


\bibitem[\protect\citeauthoryear{Ta, Le, Khoo, and Chin}{Ta
  et~al\mbox{.}}{2018}]%
        {ta:popl18}
\bibfield{author}{\bibinfo{person}{Quang-Trung Ta}, \bibinfo{person}{Ton~Chanh
  Le}, \bibinfo{person}{Siau-Cheng Khoo}, {and} \bibinfo{person}{Wei-Ngan
  Chin}.} \bibinfo{year}{2018}\natexlab{}.
\newblock \showarticletitle{Automated Lemma Synthesis in Symbolic-heap
  Separation Logic}.
\newblock \bibinfo{journal}{\emph{{PACMPL}}} \bibinfo{volume}{2},
  \bibinfo{number}{{POPL}} (\bibinfo{year}{2018}), \bibinfo{pages}{9:1--9:29}.
\newblock
\showISSN{2475-1421}


\bibitem[\protect\citeauthoryear{VCDryad}{VCDryad}{2018}]%
        {web:vcdryad}
VCDryad \bibinfo{year}{2018}\natexlab{}.
\newblock \bibinfo{title}{{Automated deductive verification framework}}.
\newblock
\newblock
\newblock
\shownote{\url{http://madhu.cs.illinois.edu/vcdryad/}.}


\bibitem[\protect\citeauthoryear{Zhu, Petri, and Jagannathan}{Zhu
  et~al\mbox{.}}{2016}]%
        {zhu:pldi16:dorder}
\bibfield{author}{\bibinfo{person}{He Zhu}, \bibinfo{person}{Gustavo Petri},
  {and} \bibinfo{person}{Suresh Jagannathan}.} \bibinfo{year}{2016}\natexlab{}.
\newblock \showarticletitle{Automatically learning shape specifications}. In
  \bibinfo{booktitle}{\emph{PLDI}}. ACM, \bibinfo{pages}{491--507}.
\newblock


\end{thebibliography}

\end{document}